\documentclass[fleqn,usenatbib]{mnras}

\usepackage[T1]{fontenc}
\usepackage{ upgreek }
\DeclareRobustCommand{\VAN}[3]{#2}
\let\VANthebibliography\thebibliography
\def\thebibliography{\DeclareRobustCommand{\VAN}[3]{##3}\VANthebibliography}

\usepackage{graphicx}	% Including figure files
\usepackage{amsmath}	% Advanced maths commands
\usepackage{amssymb}	% Extra maths symbols

\usepackage{enumitem}
\newcommand{\referee}[1]{{\color{black} #1}}
\usepackage{orcidlink}
\usepackage{algorithm}
\usepackage{dsfont}

\usepackage[noend]{algpseudocode}
\usepackage{newtxtext,newtxmath}
\makeatletter
\def\algbackskip{\hskip-\ALG@thistlm}
\makeatother

\title[HSC WL Redshift Distribution Inference]{Weak Lensing Tomographic Redshift Distribution Inference for the Hyper Suprime-Cam Subaru Strategic Program three-year shape catalogue}

\author[]{Markus~Michael~Rau$^{1,2}$\thanks{E-mail: markusmichael.rau@googlemail.com}\orcidlink{0000-0003-3709-1324}, Roohi~Dalal$^{3}$\orcidlink{0000-0002-7998-9899}, Tianqing~Zhang$^1$\orcidlink{0000-0002-5596-198X}, Xiangchong~Li$^{1}$\orcidlink{0000-0003-2880-5102}, Atsushi~J.~Nishizawa$^{6,7,8}$\orcidlink{0000-0002-6109-2397}, \newauthor  Surhud~More$^{4,5}$\orcidlink{0000-0002-2986-2371}, Rachel~Mandelbaum$^{1}$\orcidlink{0000-0003-2271-1527}, Hironao~Miyatake\orcidlink{0000-0001-7964-9766}$^{8,7,5}$, Michael A. Strauss$^{3}$\orcidlink{0000-0002-0106-7755}, Masahiro Takada$^5$\orcidlink{0000-0002-5578-6472}
\\
$^{1}$McWilliams Center for Cosmology, Department of Physics, Carnegie Mellon University, Pittsburgh, PA 15213\\
$^{2}$High Energy Physics Division, Argonne National Laboratory, Lemont, IL 60439, USA\\
$^{3}$Department of Astrophysical Sciences, Princeton University, Princeton, NJ 08544, USA\\
$^{4}$Inter University Centre for Astronomy and Astrophysics, PB 4, Ganeshkhind, Pune, 411007, IN\\
$^{5}$Kavli Institute for the Physics and Mathematics of the Universe (WPI), The University of Tokyo Institutes for Advanced Study (UTIAS), \\
The University of Tokyo, Chiba 277-8583, Japan
$^{6}$ Gifu Shotoku Gakuen University, Gifu 501-6194, Japan\\
$^{7}$ Institute for Advanced Research, Nagoya University, Nagoya 464-8602, Japan\\
$^{8}$ Kobayashi-Maskawa Institute for the Origin of Particles and the Universe (KMI), Nagoya University, Nagoya, 464-8602, Japan
}

\date{Accepted XXX. Received YYY; in original form ZZZ}

\pubyear{2021}

\begin{document}
\label{firstpage}
\pagerange{\pageref{firstpage}--\pageref{lastpage}}
\maketitle

\begin{abstract}
We present posterior sample redshift distributions for the Hyper Suprime-Cam Subaru Strategic Program Weak Lensing three-year (HSC Y3) analysis. Using the galaxies' photometry and spatial cross-correlations, we conduct a combined Bayesian Hierarchical Inference of the sample redshift distributions. The spatial cross-correlations are derived using a subsample of Luminous Red Galaxies (LRGs) with accurate redshift information available up to a photometric redshift of $z < 1.2$. We derive the photometry-based constraints using a combination of two empirical techniques calibrated on spectroscopic- and multiband photometric data that covers a spatial subset of the shear catalog. The limited spatial coverage induces a cosmic variance error budget that we include in the inference. Our cross-correlation analysis models the photometric redshift error of the LRGs to correct for systematic biases and statistical uncertainties. We demonstrate consistency between the sample redshift distributions derived using the spatial cross-correlations, the photometry, and the posterior of the combined analysis. Based on this assessment, we recommend conservative priors for sample redshift distributions of tomographic bins used in the three-year cosmological Weak Lensing analyses.
\end{abstract}


%\rachel{I would like to suggest modifying the title, short title, abstract and some of the text, for the following reason:  I think the reference to S19A is going to be quite opaque to people outside of the HSC survey.  It's an internal data release.  Moreover, we did not actually provide calibrations for all of S19A, only for the shape catalog from Li et al.  So I recommend specifying things that way everywhere (`three-year shape catalogue' or something like that) rather than in terms of S19A.} \markus{I changed the notation}

%%%%%%%%%%%%%%%%%%%%%%%%%%%%%%%%%%%%%%%%%%%%%%%%%%

%%%%%%%%%%%%%%%%% BODY OF PAPER %%%%%%%%%%%%%%%%%%
\section{Introduction}
% I added "weak lensing and structure growth" here because the papers cited are clearly only applicable there and not to e.g., supernova cosmology.
Cosmological weak lensing (WL) and structure growth analyses for the current and next generation of large area photometric surveys like the Dark Energy Survey \citep[DES; e.g.,][]{2018ApJS..239...18A}, the Kilo-Degree Survey  \citep[KiDS; e.g.,][]{2017MNRAS.465.1454H}, the Hyper Suprime-Cam  \citep[HSC; e.g.,][]{2018PASJ...70S...4A}, the Rubin Observatory Legacy Survey of Space and Time  \citep[LSST; e.g.,][]{2019ApJ...873..111I}, the Roman Space Telescope \citep[e.g.][]{2015arXiv150303757S} and Euclid \citep[e.g.][]{2011arXiv1110.3193L} depend on accurately accounting for sources of systematic bias and uncertainty \citep[e.g.][]{2018ARA&A..56..393M}. The primary cosmological probes in these campaigns are measurements of the growth of structure based on two-point statistics of galaxy and gravitational shear fields \citep[see e.g.][]{2019PASJ...71...43H, 2020PASJ...72...16H, 2021A&A...645A.105G, 2021A&A...645A.104A, 2021A&A...646A.140H, 2021A&A...646A.129J, PhysRevD.105.023515, PhysRevD.105.023514, PhysRevD.105.083528, PhysRevD.106.043520, PhysRevD.105.023520}. %\roohi{should the HSC Y1 papers (Hikage et al., Hamana et al.) be cited here too?} \markus{Added references}

%\citep[see e.g.][]{2017MNRAS.465.1454H, UitertKids, 2018PhRvD..98d3526A, 2018MNRAS.474.4894J,  2019PASJ...71...43H, 2020arXiv200715632H}.
%\rachel{Where possible, should update refs to point to the latest version (e.g., DES Y3 instead of Y1; KiDS-1000 instead of KiDS-450).}\markus{updated}

Since measurements of the broadband photometry of galaxies only allow us to extract limited redshift information, measurements of two-point statistics of density fields are typically considered in projection along the line-of-sight. The line-of-sight, or sample redshift distribution $p_{\rm samp}(z)$ enters the corresponding WL and Large-Scale Structure (LSS) theory predictions, which are used to constrain cosmological parameters using measurements of the projected density fields in a likelihood framework. In order to calibrate the credible intervals on cosmological parameters, it is important to characterize and control sources of systematic bias and uncertainty in $p_{\rm samp}(z)$ estimates \citep[see e.g.][]{10.1111/j.1365-2966.2005.09782.x, Hoyle_2018, 2018PASJ...70S...9T, 2019PASJ...71...43H, 2021A&A...647A.124H, 2020A&A...638L...1J}.

One primary science driver for photometric surveys is to constrain the dark energy equation of state parameters by measuring the distance-redshift and growth-redshift relations \citep[see e.g.,][p. 31]{dark_energy_taskforce} which both enter the WL and LSS modelling and parametrize the growth of structure and expansion history of our universe. This approach leads to degeneracies between cosmological parameters that describe the cosmic density fields, $p_{\rm samp}(z)$ parameters that enter the aforementioned line-of-sight projection kernel \citep[e.g.][]{2006ApJ...636...21M, 2010MNRAS.401.1399B}, and other modelling components such as the galaxy-dark matter bias \citep[see e.g.][]{1997MNRAS.286..115M, 2015MNRAS.448.1389C, 2016MNRAS.459.3203C, 2018A&A...613A..15S, 2018MNRAS.473.1667P, 2020PhRvD.102h3520S, 10.1093/mnras/stac3630} \referee{and intrinsic alignments \citep{PhysRevD.105.023514, 2022PhRvD.105b3515S, PhysRevD.105.083529}.} Parameters that describe the sample redshift distribution for samples of galaxies, can therefore exhibit a degeneracy with cosmological or astrophysical parameters.
Inaccuracies in the distance (or redshift) measurements of ensembles of galaxies are therefore important \referee{for} modelling systematics in these surveys.

The two main sources of information available to constrain redshifts of individual galaxies as well as samples of galaxies are measurements of their photometry and spatial clustering. Methods that exploit photometric information \citep[for a recent review, see][]{2019NatAs...3..212S, 2022ARA&A..60..363N} can be broadly categorized into two classes.  Empirical methods \citep{2003LNCS.2859..226T, 2004PASP..116..345C, 2010ApJ...715..823G, 2013MNRAS.432.1483C,  2015MNRAS.449.1043B, 2015MNRAS.452.3710R, 2016A&C....16...34H} utilize calibration data to directly learn a mapping from the measured photometry to the redshift of galaxies given a spectroscopic survey.  Template fitting methods \citep[e.g.,][]{1999MNRAS.310..540A, 2000ApJ...536..571B,2006A&A...457..841I, 2006MNRAS.372..565F, 2015MNRAS.451.1848G, 2016MNRAS.460.4258L, 2020arXiv200712178M} use a forward model that constrains the redshift of galaxies using a likelihood of the `reproduced' galaxy flux, given a model for the galaxy Spectral Energy Distribution (SED) and other parameters of interest.

Both of these approaches lead to consistent estimators if their underlying assumptions are met and a correct statistical estimator is constructed. However, in real data, incorrectly modelled selection functions and modelling uncertainties can lead to significant model misspecification. A particular example are selection functions in spectroscopic datasets used for redshift calibration \referee{\citep{2017ApJ...841..111M, 2019ApJ...877...81M, 2020MNRAS.496.4769H}}, due to the impractically long exposure times required to spectroscopically observe color-complete samples at faint magnitudes \citep[see e.g.][]{10.1093/mnras/stu1424, 2015APh....63...81N}. One goal of this paper is to discuss and discern the assumptions made in various $p_{\rm samp}(z)$ inference methodologies by discussing them in a unified likelihood framework.

As mentioned, a second method to constrain $p_{\rm samp}(z)$ are spatial cross-correlations between photometric and spectroscopic samples \citep[e.g.][]{2008ApJ...684...88N, 2013arXiv1303.4722M, 2013MNRAS.433.2857M, 2016MNRAS.462.1683S, 10.1093/mnras/stx691, Morrison2016, 2017arXiv171002517D, 2018MNRAS.477.1664G, 2020A&A...642A.200V, 2021A&A...647A.124H}. Since the photometric and spectroscopic samples trace the same underlying dark-matter field, the amplitudes of the two-point function measured between spectroscopic samples (binned in redshift) and the full photometric sample (with no accurate redshift information) can constrain the sample redshift distribution of the full photometric sample $p_{\rm samp}(z)$. Redshift-dependent galaxy-dark matter bias of the photometric and spectroscopic samples, cosmic magnification effects  \citep[see e.g.][]{2005ApJ...633..589S} and the redshift evolution of the underlying dark-matter density field affect the aforementioned relative redshift bin heights.

While it is a challenge to correct for these degenerate effects, cross-correlations are one of the most important techniques for $p_{\rm samp}(z)$ calibration today. We note that two point statistics from e.g., weak lensing \citep[e.g.][]{2013MNRAS.431.1547B, 2020arXiv201207707S}, or shear-ratios \citep[e.g.][]{2019MNRAS.487.1363P, 2021A&A...645A.105G, 2021arXiv210513542S, PhysRevD.105.083529} can also be used in the context of redshift estimation. 

However since weak lensing in particular is considered one of the most promising methods to constrain dark energy, photometric redshift estimation is treated in our analysis as a systematic that enters the theoretical modelling of a separate `cosmological' likelihood rather than using WL statistics as a redshift estimation technique. Recently, the question of how to integrate redshift uncertainty into a likelihood of two point statistics has been considered \citep{2017MNRAS.466.3558M, 2018arXiv180202581H}, especially in the context of how to combine template fitting and cross correlation measurements \citep{2019arXiv191007127A, 2019MNRAS.483.2801S, 2019MNRAS.483.2487J, 2020MNRAS.491.4768R,2021MNRAS.505.4249M, 2022MNRAS.510.1223G, 2020arXiv201212826C, 2022MNRAS.509.4886R, 2022arXiv220610169Z}. In \citet{2022MNRAS.509.4886R}, we developed a Bayesian hierarchical inference framework that self-consistently combines information from both cross-correlation redshift estimation and photometry, specifically discussing aspects of regularization and probability calibration. \referee{\citet{2022MNRAS.509.4886R} validates the basic aspects of our presented methodology using mock data where well-controlled sources of systematics are modelled. While the usage of simulated mock data necessarily has limitations, we performed this analysis with the greatest possible realism in mind. We found that a hierarchical modelling approach similar to the one presented in this paper can indeed reach the level of accuracy necessary for LSST, as measured using common performance metrics. }

\referee{This paper presents the sample redshift inference methodology for the HSC Y3 cosmological weak lensing analysis, which consists of two cosmic shear analyses \citep{2023arXiv230400701D, 2023arXiv230400702L} in four tomographic bins and a 3x2pt analysis \citep{2023arXiv230400703M, 2023arXiv230400704M, 2023arXiv230400705S} which uses one tomographic bin. This paper presents our inference methodology in the context of the cosmic shear analyses, where it was used as the default method for redshift inference.}  % Note previous sentence implied 3x2pt used 4 tomo bins; rearranged to fix
Tomography refers here to binning the shear catalog along the redshift dimension, using a predictor for redshift. While the separation of these tomographic samples in redshift is typically not perfect, i.e., the sample redshift distributions of adjacent tomographic bins will overlap, auto- and cross-correlations estimated on the tomographic samples will have more information about the redshift evolution of the growth of structure than the two-point function estimated on the unbinned sample. We utilize 5 band photometry in the $grizy$ filter set to infer the tomographic sample redshift distributions (tomographic $p_{\rm samp}(z)$). We apply our methodology to the Hyper Suprime-Cam three-year shape catalog\footnote{Data observed through 2019.} dataset (HSC Y3), and derive and recommend prior distributions over a $p_{\rm samp}(z)$ parameterization that can be used in the subsequent cosmological weak lensing analyses. 

This work presents a significant update to the HSC sample redshift inference methodology developed for the first year (HSC Y1) analyses presented in \citet{2019PASJ...71...43H} and \citet{2020PASJ...72...16H}. This is vital, since the increased area of the shear catalog from $136.9 \, {\rm deg}^2$ (HSC Y1) to $433.5 \, {\rm deg}^2$ (HSC Y3) implies that our redshift calibration accuracy has to significantly improve to prevent systematic biases or uncertainties in cosmological parameters from dominating over the statistical uncertainties. 

\section{Motivation}
\label{sec:impact_s16a_cosmology}
The HSC Y1 sample redshift distribution calibration described in \citet{2019PASJ...71...43H} and applied in the context of the Y3 cosmic shear analysis in that work and in \citet{2020PASJ...72...16H} estimates the sample redshift distributions in tomographic bins by reweighting COSMOS2015 \citep{2016ApJS..224...24L, 2006A&A...457..841I} galaxies in color space. The quantification of uncertainty includes a systematic error budget derived by comparing the reweighted sample redshift distribution with the sample redshift distribution estimators obtained from a set of 7 independent methods. The HSC Y1 analyses used uncertainties in the means of the tomographic redshift distributions as parameters to marginalize over photometric redshift uncertainty.

The forthcoming HSC Y3 analyses also include a systematic error budget based on a comparison of models, but presents a significantly updated framework for sample redshift inference that includes a treatment of cosmic variance as well as a cross-correlation calibration of sample redshift distributions based on a sample of Luminous Red Galaxies \citep[][]{2014MNRAS.444..147O, 2018PASJ...70S..20O, 2018PASJ...70S..26O, 2021ApJ...922...23I} selected using the Cluster finding algorithm based on Multi-band Identification of Red-sequence gAlaxies (CAMIRA). We will abbreviate this sample as `CAMIRA LRG' in the following. The inclusion of a cross-correlation data vector into the inference of the sample redshift distribution $p_{\rm samp}(z)$ is arguably the most significant improvement over the Y1 analyses, as it allows us to independently test the quality of the estimated $p_{\rm samp}(z)$ in the tomographic bins.

\begin{figure}
    \centering
    \includegraphics[scale=0.4]{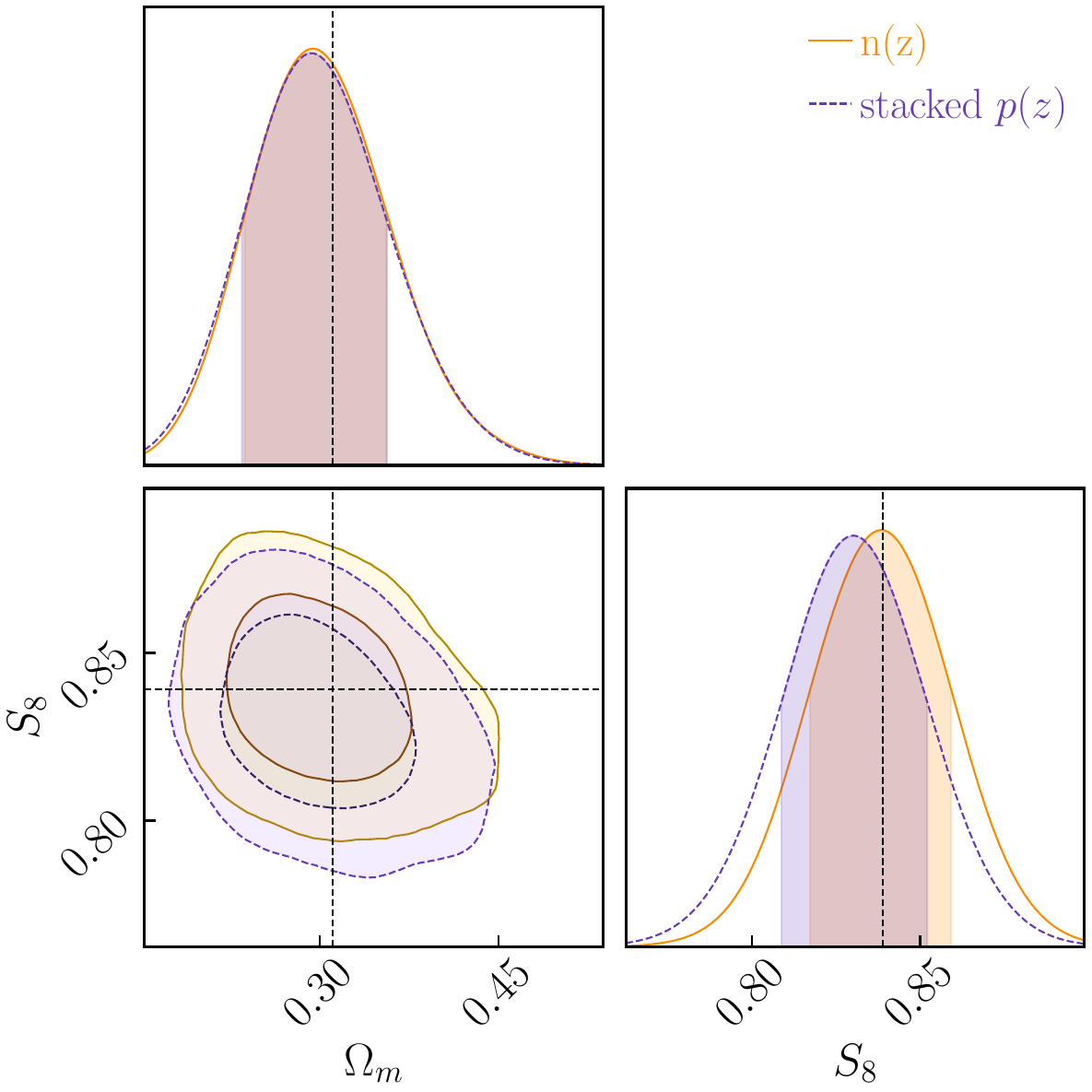}
    \caption{ Forecast of the impact and importance of using an updated $p_{\rm samp}(z)$ inference methodology on cosmological inference from the predicted Y3 data vector and covariance matrix. The purple contour uses the `stacked' redshift distribution for the Y1 galaxy catalog, while the orange contour uses the Y1 redshift distribution inferred from an analysis similar to this work. The change in redshift distribution causes a $0.5\sigma$ shift in the $S_8$ constraints, which is significant for the upcoming Y3 cosmic shear analyses.  }
    \label{fig:full_y3_takahashi}
\end{figure}
% As described in \S~\ref{sec:discussion}, we have updated our photometric redshift inference methodology significantly from that in \citet{2020PASJ...72...16H}. The goal of this appendix is to
We refer to the remainder of the paper for an explanation of the HSC Y3 redshift inference methodology. However we would like to motivate the effect that these significant changes have on our redshift calibration using a forecast, which is based on a mock Y3 weak lensing cosmological analysis. We perform a mock analysis of a synthetic data vector with a redshift distribution inferred for the HSC Y1 shape catalog, using a similar analysis to the one presented in this work, and compare it with an analysis using the simple `stacked' redshift distribution from \citet{2020PASJ...72...16H}. The main difference from the methodology described in the rest of this work is the usage of the Dirichlet distribution, as well as the usage of a model combination scheme described in Rau et al (in prep.) that accounts for the model uncertainty across the several different photometric redshift codes applied to the HSC Y1 dataset and described in  \citet{2020PASJ...72...16H}. 

%\rachel{The text below is somewhat unclear regarding exactly which redshift distribution was used to generate the mock data vector (it is just clear that two different distributions were used for the analysis).}\tianqing{I have added a sentence address that the data vector is generated by the updated redshift distribution. }
The sample redshift posteriors and the inference scheme employed to marginalize over the uncertainty in those parameters are both described in \citet{2022arXiv220610169Z}. The sampling is based on using the mean of the tomographic redshift distributions as the main parameter over which we marginalize (referred to as the `shift model' in the following). The cosmological parameter inference is performed using the multinest method with 500 live points. \referee{We consider 9 cosmological and 9 astrophysical parameters and 4 parameters within the shift model. The simulated data vector includes noise based on the scaled HSC first year covariance as described in \citet{2022arXiv220610169Z}.} Both contours shown in Fig.~\ref{fig:full_y3_takahashi}  use the shift model to marginalize over the $p_{\rm samp}(z)$ uncertainties, where our prior on the tomographic $p_{\rm samp}(z)$ is generated using the mean redshifts of 1000 samples of sample redshift posterior generated by the updated methodology. The prior on the mean redshift of the stacked redshift distribution follows \citet{2020PASJ...72...16H}. We generate an approximation to the Y3 covariance by dividing the Y1 covariance by 3, which approximately accounts for the increase in area from Y1 to Y3 while ignoring changes in the contiguity of the survey footprint. Fig.~\ref{fig:full_y3_takahashi} compares the posteriors in the $\Omega_m-S_8$ plane. 

We note a 0.5$\sigma$ shift in $S_8$, which shows that the updated analysis would predict a higher $S_8$ value. Note that the synthetic data vector is generated with the updated redshift distribution, so the analysis with that redshift distribution recovers the true cosmological parameters. This figure illustrates the importance of $p_{\rm samp}(z)$ calibration and in particular of a joint $p_{\rm samp}(z)$ analysis that includes complementary data sources and analysis techniques.

\section{Data}
\label{sec:data}
The following sections describe the datasets and catalogs that we use in this work. Specifically, we consider three datasets that are relevant at different stages of the analysis. \S~\ref{subsec:hsc_data} describes the photometric data included in the HSC shear catalog, \S~\ref{subsec:camira_data} the catalog of Luminous Red Galaxies \citep[LRGs, ][]{2014MNRAS.444..147O, 2018PASJ...70S..20O, 2018PASJ...70S..26O, 2021ApJ...922...23I} that we will use for our cross-correlation analysis and \S~\ref{subsec:spec_reference} a matched catalog between the photometric data and spatially overlapping spectroscopic surveys. We will abbreviate the photometric data included in the HSC shear catalog as 'HSC phot`, the catalog of Luminous Red Galaxies as 'CAMIRA LRG` and the matched catalog as 'specXphot`.

\subsection{HSC Y3 Shape Catalog}
\label{subsec:hsc_data}
The Hyper Suprime-Cam survey, which is part of the Subaru Strategic Program (SSP), is an optical imaging survey carried out using the Hyper Suprime-Cam (HSC, \citealt*{2018PASJ...70S...1M}), a wide field camera with 1.77 ${\rm deg}^2$ field of view installed on the 8.2 meter Subaru telescope.
% The total survey footprint is split into a Wide (4200 ${\rm deg^2}$) \xlnt{1200, is the number for the final year}, Deep ($\sim$25 ${\rm deg}^2$) and Ultra-Deep field ($\sim$3.5 ${\rm deg}^2$). 
The shear catalog we use in this work, as part of the year-3 analysis, consists of $417$~deg$^2$\footnote{We remove a $20$~deg$^2$ region that failed the cosmic shear B-mode test (see Zhang et al. in prep.).} of wide-field optical galaxy photometry in $g$$r$$i$$z$$y$ with a $5 \sigma$ limiting magnitude of $r \approx 26$. 
We refer the reader to \citet{2018PASJ...70S...4A} and \citet{2022PASJ...74..247A} for a more detailed overview of the design of the HSC survey. The catalogs from this internal data release along with the shape catalog and their calibrations are expected to be made public as part of a future incremental update to PDR3 \citep{2022PASJ...74..247A} after the cosmological analyses are finished.

\begin{figure}
    \centering
    \includegraphics[scale=0.6]{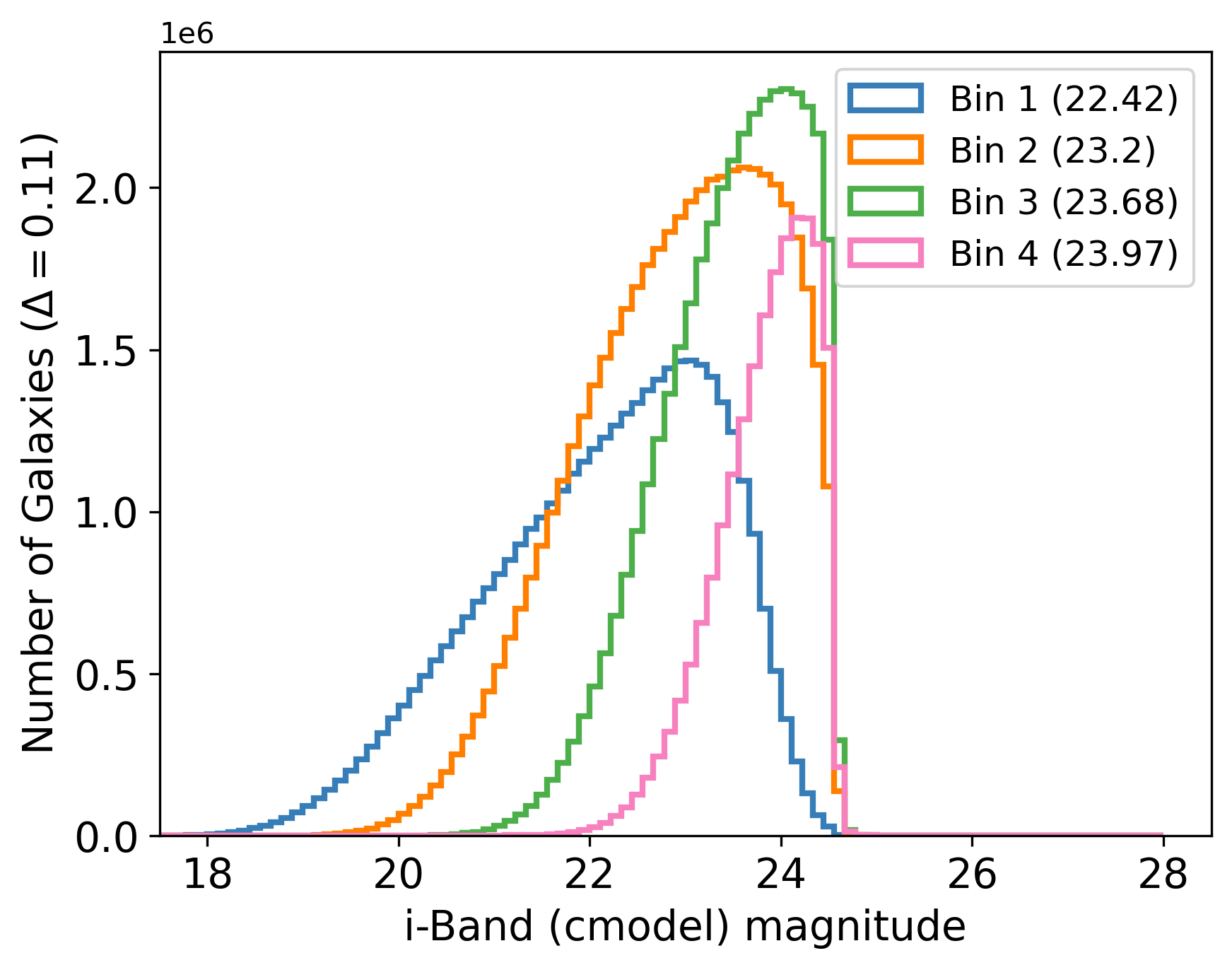}
    \caption{Distribution of $i$-band cmodel magnitudes for the four tomographic bins. We show the $i$-band cmodel magnitudes on the horizontal axis and the number of galaxies on the y-axis. The median magnitudes are shown in the legend, the magnitude bin size is $\Delta = 0.11$. }
    \label{fig:magnitude_distribution}
\end{figure}

Fig.~\ref{fig:magnitude_distribution} plots the cmodel\footnote{The SDSS CModel magnitude \citep{2001ASPC..238..269L, 2004AJ....128..502A} algorithm fits a galaxy using elliptical models with both an exponential profile and a de Vaucouleurs profile. The derived CModel flux is approximately a linear interpolation between exponential and de Vaucouleurs models. We refer to \citet{10.1093/pasj/psx126} for more details.} magnitude distribution in the $i$-band for the four tomographic bins. The tomographic bins (`Bin 1', `Bin 2', `Bin 3', `Bin 4') are selected using a procedure described in \S~\ref{subsec:sample_selection} to have approximately the redshift ranges of (0.3, 0.6], (0.6, 0.9], (0.9, 1.2], and (1.2, 1.5]. 

We see that all four tomographic bins extend to magnitudes fainter than $24$ in
the $i$-band, where the majority of galaxies have a magnitude around that
value. Bins~1--4 contain 24, 33, 28 and 15 per cent of the galaxies,
respectively, and the raw (effective) galaxy number densities are
$3.92~(3.77)$, $5.63~(5.07)$, $4.68~(4.00)$ and
$2.60~(2.12)~\mathrm{arcmin}^{-2}$.
Since we present this analysis in the context of the upcoming cosmic shear analysis for HSC Y3, we apply our methodology to galaxies contained in the shear catalog that has a magnitude limit of 24.5. We, therefore, need to include all of the lensing selection criteria and lensing weights throughout the analysis. Lensing weights are inverse variance weights derived in the construction of the galaxy shape estimate. For a description of the methodology to derive these selection criteria and lensing weights, we refer to \citet{2022PASJ...74..421L}. In the following, we will refer to the shear catalog as `HSC phot'. 

\subsection{CAMIRA LRG Sample}
\label{subsec:camira_data}
\begin{figure}
    \centering
    \includegraphics[scale=0.55]{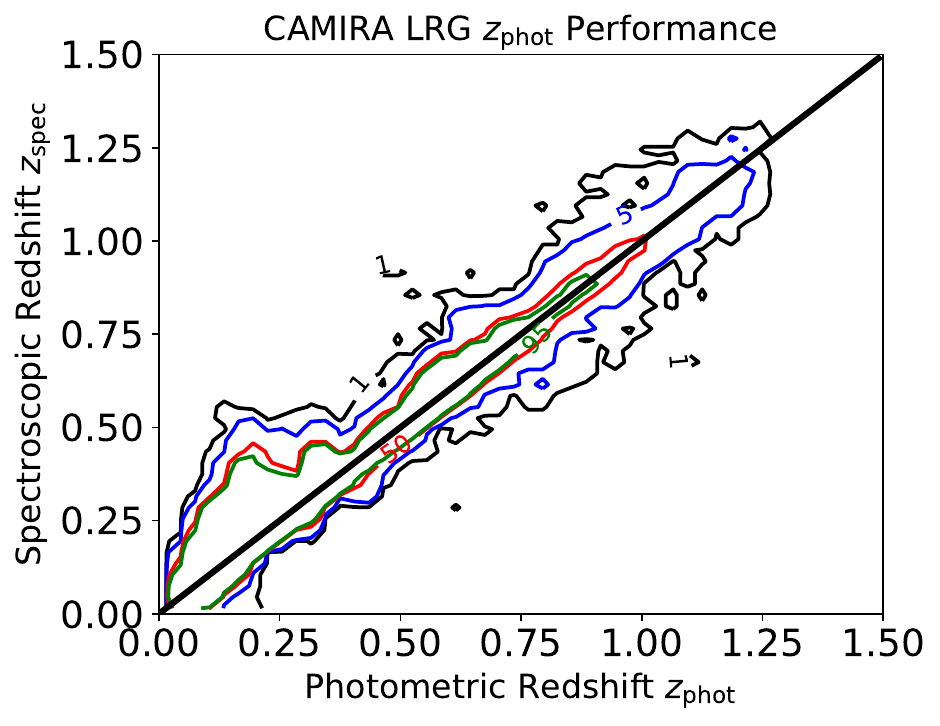}
    \caption{ Photometric vs.\ spectroscopic redshift for CAMIRA LRG galaxies with spectroscopic redshifts. The dashed black line denotes perfect photometric redshift prediction. There is a small population (0.02\%) of redshift outliers at $z_{\rm spec} > 5$ 
    that we do not show here. The contour line annotations indicate the corresponding probability density values in per-cent.
    }
    \label{fig:zphot_vs_zspec_comparison}
\end{figure}

\begin{figure}
    \centering
    \includegraphics[scale=0.6]{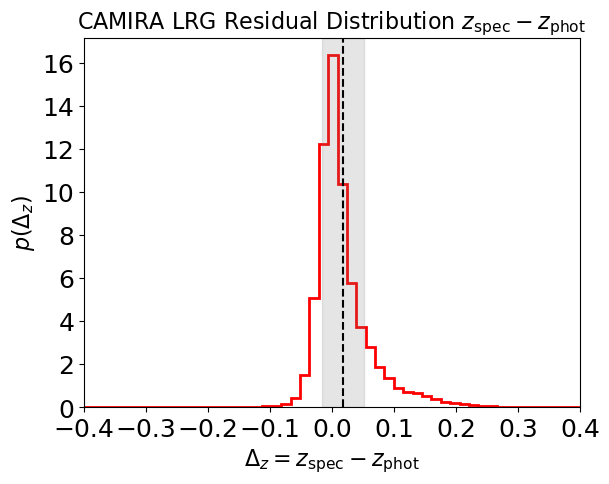}
    \caption{ Distribution of photometric redshift residuals of $z_{\rm spec} - z_{\rm PhotZ}$. The black dashed vertical line denotes the mean, while the grey contours show the range between the 16th and 84th percentiles (selected to resemble a `Gaussianized' $1\sigma$ interval).  
    }
    \label{fig:residual_camira_lrg}
\end{figure}

The CAMIRA Luminous-Red-Galaxy (LRG) sample\footnote{\url{https://github.com/oguri/cluster_catalogs/tree/main/hsc_s20a_camira}, last accessed 10/06/2022} contains Luminous Red Galaxies selected using the CAMIRA algorithm (Cluster finding Algorithm based on Multi-band Identification of Red-sequence Galaxies; \citealt*{2014MNRAS.444..147O, 2018PASJ...70S..26O, 2021ApJ...922...23I}). CAMIRA identifies LRGs as red-sequence galaxies based on their photometry and their consistency with the expected colors from stellar population synthesis models. The LRG sample has a limited redshift range of $z < 1.2$ and the redshifts of these LRGs are subject to photometric redshift error\footnote{Photometric Redshifts for LRGs are often derived using SED fitting techniques and have significantly better redshift accuracy compared with the full photometric sample.}. In this work we use the CAMIRA LRG sample as a reference catalog for spatial cross-correlations with galaxy samples from HSC phot. This will allow us to construct a likelihood that constrains the $p_{\rm samp}(z)$. Since the LRG galaxy population provides a photometric sample with good redshift quality and well understood clustering properties, it is the ideal reference sample for cross-correlation studies. However, as we will describe in \S~\ref{subsec:joint_inference_camira}, we need to marginalize over the photometric redshift error of the LRGs. This requires a model for the photometric redshift error of the CAMIRA LRG galaxies, which we detail there. The photometric redshift error model is calibrated using the corresponding LRG subsample of the full specXphot reference sample described in \S~\ref{subsec:spec_reference}.
Fig.~\ref{fig:zphot_vs_zspec_comparison} shows the photometric redshift of the CAMIRA LRGs against the spectroscopic redshifts of the aforementioned specXphot reference subsample as a contour plot. We see that especially around ($z_{\rm spec} \approx 0.4$/$z_{\rm phot} \approx 0.2$), a well-known redshift region where the 4000{\AA} break crosses between the $g$ and the $r$ filters, the photometric redshift of the CAMIRA LRG galaxies shows a mean bias in the contour lines. Although, we identify a small number of outlier galaxies with $z_{\rm spec} > 5$. This population consists of 0.02\% of the full CAMIRA LRG specXphot reference sample; the contamination is small and we leave a further investigation of the outlier population for future work. The bias at low photometric redshift is also apparent in the right tail of Fig.~\ref{fig:residual_camira_lrg}, which shows
a histogram of the residual redshift error $z_{\rm spec} - z_{\rm Phot}$.
% RM: note, it said PhotZ but this is inconsistent with the plot labeling
The black dashed vertical line shows the mean residual redshift error (0.018), while the grey region visualizes the range between the 16th (-0.017) and 84th (0.052) percentiles %\surhud{Shouldn't this be 16 and 84?}
(equivalent to the `Gaussian' $\pm 1\sigma$ intervals).

\subsection{Spectroscopic Reference Samples}
\label{subsec:spec_reference}
\begin{figure}
    \centering
    \includegraphics[scale=0.7]{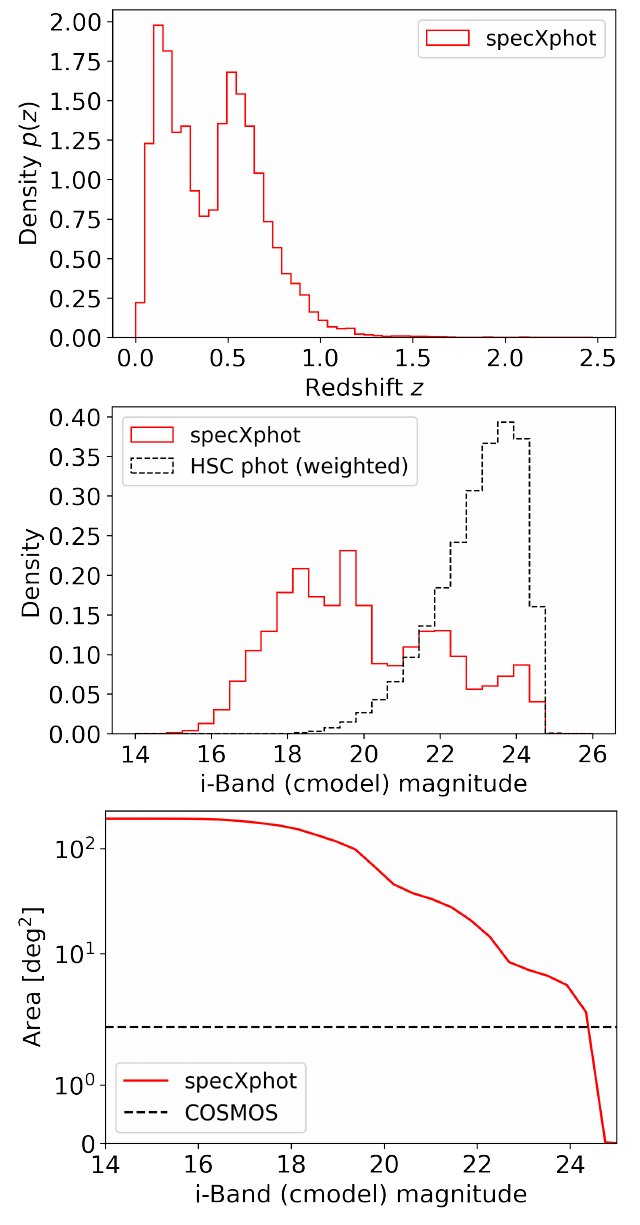}
    \caption{ %\markus{
    Illustration of the spatial coverage and the distribution of galaxies as a function of $i$-band magnitude for the specXphot Calibration dataset used for $p_{\rm indiv}(z)$ estimation. \textit{Top:} Spectroscopic redshift distribution of the specXphot calibration sample. The histogram is normalized to integrate to unity. \textit{Middle:} Distribution of galaxies in $i$-band magnitude for the specXphot Calibration dataset (red solid) and the HSC phot dataset (black dashed) including lensing weights. \textit{Lower:} Area in square degrees covered by the specXphot dataset as a function of $i$-band magnitude. The vertical axis, plotted on the symmetrical log scale, shows the total area covered by all galaxies with $i$-band magnitude brighter  
    than the value shown on the horizontal axis. The dashed horizontal line shows the area covered by the COSMOS2015 dataset that dominates the specXphot dataset at the faint end. 
    }
    \label{fig:spatial_selection_function}
\end{figure}
This section gives an overview of the spectroscopic reference samples that are available to match against HSC phot to generate the `specXphot' calibration sample. We will concentrate on the aspects that are relevant for this work and refer to \citet{2018PASJ...70S...9T} for a more detailed description of the reference samples and the selection criteria used to generate them. %\rachel{I thought they were described in Tanaka et al, not Aihara et al?  (question applies to this reference and the later ones to Aihara et al in this section)} \markus{Corrected}

\referee{The reference sample \citep{Nishizawa:2020}
%(see Nishizawa et al. in prep. \footnote{\url{https://hsc-release.mtk.nao.ac.jp/doc/wp-content/uploads/2022/08/pdr3_photoz.pdf} (Accessed Oct. 2022)})
} 
is assembled from the following sources:  zCOSMOS DR3 \citep{2009ApJS..184..218L}, \referee{zCOSMOS faint \citep{2009ApJS..184..218L} including private spectroscopic data}\footnote{Mara Salvato private communication.}, COSMOS2015 \citep{2016ApJS..224...24L},
UDSz \citep{2013MNRAS.433..194B,2013MNRAS.428.1088M},
3D-HST \citep{2014ApJS..214...24S,2016ApJS..225...27M},
FMOS-COSMOS \citep{2015ApJS..220...12S},
VVDS \citep{2013A&A...559A..14L},
VIPERS PDR1 \citep{2014A&A...562A..23G},
SDSS DR12 \citep{2015ApJS..219...12A},
GAMA DR2 \citep{2015MNRAS.452.2087L},
WiggleZ DR1 \citep{2010MNRAS.401.1429D},
DEEP2 DR4 \citep{2003SPIE.4834..161D,2013ApJS..208....5N}, \referee{ VANDELS DR2 \citep{2018A&A...616A.174P}, C3R2 \citep{2017ApJ...841..111M, 2019ApJ...877...81M}}, and
PRIMUS DR1 \citep{2011ApJ...743...46C,2013ApJ...767..118C}. The spectroscopic redshift measurements are extracted from both high-quality spectroscopic measurements ($\approx$ 170,000 galaxies) and lower resolution prism spectroscopy ($\approx$ 37,000 galaxies). In addition, \citet{2018PASJ...70S...9T} also include 170,000 Cosmos2015 multiband photometric redshifts and a sample of privately obtained spectroscopic redshifts (Mara Salvato private communication).

\citet{2018PASJ...70S...9T} homogenize the catalog to ensure approximately uniform data quality. This is done by imposing cuts on the quality flags in the respective source catalogs. The galaxies are then matched to HSC phot (see \S~\ref{subsec:hsc_data}) to create the specXphot reference sample. This catalog contains both the photometric measurements in HSC phot and the spectroscopic redshift estimates from the listed sources.

We will utilize this dataset as a reference sample to calibrate and train photometric redshift estimates. While the selection cuts imposed by \citet{2018PASJ...70S...9T} are designed to minimize the impact of color-redshift incompleteness on photometric redshift estimates trained on the specXphot calibration sample, we still have to consider the spatial selection function due to the much smaller survey footprint of the specXphot sample in relation to HSC phot. Furthermore, residual selection function induced systematics will likely remain, which motivates our usage of cross-correlations for redshift calibration.

To give an overview of this dataset, Fig.~\ref{fig:spatial_selection_function} shows the normalized spectroscopic redshift distribution of the specXphot sample (upper panel), the histogram of the $i$-band magnitude (middle panel), and the spatial area covered by the specXphot calibration catalog up to (i.e., fainter than) the magnitude limit plotted on the horizontal axis (lower panel). The middle panel shows that the specXphot calibration catalog covers the magnitude range of the HSC phot sample (black dashed histogram). We generate the lower panel by adding up the area as a function of $i$-band magnitude covered by the specXphot calibration catalog using a healpix pixelization \citep{2005ApJ...622..759G} with ${\rm NSIDE} = 1024$. The black dashed horizontal line shows the size of the COSMOS2015 calibration field ($\approx 2 \, {\rm deg}^2$) that dominates the data at the faint end. It represents the lower limit on the HSC Y3 area, for which we have available calibration data. This lower limit will be used in \S~\ref{subsec:sample_pz_PhotZ} to derive a conservative assessment of the cosmic variance error budget in our $p_{\rm samp}(z)$ inference methodology.

\section{The photometric redshift problem}
\label{sec:photometric_redshift_problem}
The $p_{\rm samp}(z)$ of galaxies is a vital component in the modelling of projected density fields in weak gravitational lensing and large-scale structure. This 1-point density distribution along the line-of-sight enters the projection kernel in the modelling of these probes. 
In this section, we summarize the foundational methodology for estimating the redshift distributions of galaxy ensembles (`$p_{\rm samp}(z)$ inference', hereafter). 

There are two main approaches to the photometric redshift problem. The `forward-modelling' approach models the data generating process\footnote{The `data generating process' refers to the procedure of drawing galaxy properties from population distributions like the sample redshift distribution and mapping these quantities to measured observables, like e.g. the photometry, via a likelihood (or sampling distribution). }  and treats the $p_{\rm samp}(z)$ as the prior on the redshift of individual galaxies. We note that `traditional' approaches like SED fitting would also fall under this category. The alternative `conditional density estimation' approach, constructs a direct probabilistic mapping between the photometry of galaxies and their redshift. For HSC, we consider both methodologies, and therefore describe $p_{\rm samp}(z)$ inference in both scenarios in the following two subsections. We note however that the models that we select for our final inference (`DNNz' and `DEMPz', see \S~\ref{subsection:indiv_galax_z}) are both conditional density estimation techniques. We still describe both methodologies in detail for completeness. 

Throughout this paper we parameterize the $p_{\rm samp}(z)$ using a histogram with height parameters $\boldsymbol{\phi_\text{nz}}$ for $N_{\rm bins}$ histogram bins as
\begin{equation}
    p_{\rm samp}(z) = \sum_{i=1}^{N_{\rm bins}} \phi_{\rm nz, i} \mathds{1}_i(z) \, ,
    \label{eq:parametrization_pz}
\end{equation}
where $\mathds{1}_i$ denotes the `indicator' function for a given histogram bin $i$. The indicator function $\mathds{1}(z)$ is unity if $z$ falls in the histogram bin, and zero otherwise.  We note that instead of a histogram parametrization one could also consider a kernel ansatz using, e.g., a Gaussian kernel. This could have advantages, because we could consider a continuous approximation with (potentially) fewer parameters. However this is not expected to be a vital reduction in approximation error. In the current analysis we decided to use the histogram, a flexible parametrization that does not necessitate the development of a specialized model for the $p_{\rm samp}(z)$.   
In the following subsections we will describe two methodologies to infer sample redshift distributions $p_{\rm samp}(z)$.

We want to briefly (and somewhat colloquially) comment on the different interpretation of $p_{\rm samp}(z)$ in both contexts. Both techniques formulate a likelihood for the parameters $\boldsymbol{\phi}$. The likelihood formulated in \S~\ref{subsec:forward_modelling} describes a sampling distribution over the observed flux. The approach \S~\ref{subsec:conddens} describes a sampling distribution over parameters of a density estimate constructed using conditional density estimates that map directly from observed photometry to galaxy redshift. We highlight that it is important to distinguish both approaches and continue with a detailed description of each in the following subsections.      

\subsection{Forward-Modelling Approach}
\label{subsec:forward_modelling}
The goal of the forward modelling approach in general and SED modelling in particular is to formulate a statistical procedure that hierarchically models the relation between ensemble distributions of quantities of interest like galaxy redshift, type or stellar mass, the corresponding properties of individual galaxies and observables like photometry.

In a simplified model (focussing on the redshift $z$ as the quantity of interest) we can formulate this as \citep[e.g.][]{2016MNRAS.460.4258L, 2020arXiv200712178M, 2022MNRAS.509.4886R}
\begin{equation}
    p(\mathbf{\hat{F}} | \boldsymbol{\phi_\text{nz}}, \boldsymbol{\Omega}) = \prod_{i = 1}^{N_{\rm gal}} \int \mathrm{d}z_i \, \omega_i \, p(\mathbf{f}_i | z_i, \boldsymbol{\Omega}) p(z_i | \boldsymbol{\phi_\text{nz}}, \boldsymbol{\Omega}) \, .
    \label{eq:main_text_forward_modelling}
\end{equation}
Here, $\mathbf{\hat{F}}$ denotes the set of fluxes of all $N_{\rm gal}$ galaxies in the sample, $\mathbf{f}_i\,(z_i)$ denotes the flux in a filter set (redshift) of the individual galaxy with index $i$, and $\boldsymbol{\Omega}$ denotes a set of auxiliary parameters that describe other galaxy properties such as galaxy type or stellar mass. The factor $\omega_i$ denotes the lensing weight for galaxy $i$. We note that bold symbols denote vector quantities. Eq.~\eqref{eq:main_text_forward_modelling} assumes that the flux and redshift of each galaxy are drawn independently of any other. To simplify the notation we will implicitly assume conditioning on $\boldsymbol{\Omega}$, but omit it from the notation in the following discussion. \referee{Effects like blending \citep{2022MNRAS.509.3371M, 2023A&A...670A.100L} break the aforementioned assumption of independence of the galaxy flux measurements. This requires either the formulation of a joint flux likelihood of sets of galaxies or a reformulation of the likelihood on the pixel level to facilitate a joint inference with photometry and shear. We do not expect this approximation to dominate the error budget for this analysis and refer to future work. Also, note that \citet{2022PASJ...74..421L} explored the connection between redshift and shear calibration in the context of simulations devised to explore blending effects for HSC survey data, and have already folded this effect into our understanding of redshift-dependent shear calibration.}  

We identify the term $p(\mathbf{f}_i | z_i, \boldsymbol{\Omega})$ in Eq.~\eqref{eq:main_text_forward_modelling} as the likelihood of the observed individual galaxy flux given redshift, and the term $p(z_i | \boldsymbol{\phi_\text{nz}}, \boldsymbol{\Omega})$ as the prior distribution of the galaxy redshifts given the parameters that describe the sample redshift distribution (see Eq.~\ref{eq:parametrization_pz} for the definition of these parameters). This specifies a forward model, where the individual galaxy redshifts $z_i$ are first `drawn' from the sample redshift distribution, denoted by the prior $p(z_i | \boldsymbol{\phi_\text{nz}}, \boldsymbol{\Omega})$. The likelihood then relates the drawn galaxy redshifts $z_i$ to the observed galaxy fluxes $\mathbf{f}_i$ via the likelihood function $p(\mathbf{f}_i | z_i, \boldsymbol{\Omega})$. We note, that the sample redshift distribution $p_{\rm samp}(z)$ is here conditional on both the parameters $\boldsymbol{\phi_\text{nz}}$ that are used to construct the distribution, as well as auxillary parameters $\boldsymbol{\Omega}$ that describe other quantities of interest.

In the following we present a toy model that illustrates some aspects of the forward model formulation in a more concise manner. We also refer to \citet{meister2009deconvolution}, \citet{2022MNRAS.509.4886R} and \citet{2005MNRAS.359..237P} for similar introductions.
Simplifying the problem and notation we can relate Eq.~\eqref{eq:main_text_forward_modelling} to the linear model
\begin{equation}
    \boldsymbol{\phi_\text{nz}}_{\rm noisy} = \boldsymbol{K} \cdot \boldsymbol{\phi_\text{nz}}_{\rm true}
    \label{eq:simple_linear}
\end{equation}
by identifying $p(\mathbf{\hat{F}} | \boldsymbol{\phi_\text{nz}}, \boldsymbol{\Omega})$ with a `smeared-out' and observed vector  $\boldsymbol{\phi_\text{nz}}_{\rm noisy}$, the set of likelihoods $\{p(\mathbf{f}_i | z_i, \boldsymbol{\Omega}) \, | \, 0 < i < N_{\rm gal} \}$ with the matrix $\boldsymbol{K}$ and the sample redshift distribution $p(z_i | \boldsymbol{\phi_\text{nz}}, \boldsymbol{\Omega})$ with a noiseless, or `true', vector $\boldsymbol{\phi_\text{nz}}_{\rm true}$.

Thus to recover $\boldsymbol{\phi_\text{nz}}_{\rm true}$ we need to invert the matrix $\boldsymbol{K}$, which can be very sensitive to small variations in $\boldsymbol{\phi_\text{nz}}_{\rm noisy}$ or the matrix $\boldsymbol{K}$. The former could be caused, for example, by the photometric noise, the latter by model error in the forward model. The sensitivity of the linear model on these variations depends on the condition number of $\boldsymbol{K}$, that will in turn depend on the resolution of the reconstruction, i.e., the histogram width in our parametrization. The forward modelling approach therefore treats $p_{\rm samp}(z)$ inference as an inverse problem whose solution is critically dependent on accurate modelling of the individual galaxy likelihoods and the regularization strategies that we impose. The likelihood modelling should also include how galaxies are selected into tomographic bins and other selection functions. 

Typically one needs to `regularize' this inverse problem. Regularization techniques reduce the noise in the reconstructed $p_{\rm samp}(z)$ by adding constraints to its shape. Ideally this information is not chosen arbitrarily, but rather results from data driven constraints (e.g., a cross-correlation data vector that is included into the inference). We refer to a more detailed discussion on regularization and its methodological challenges in our previous work \citep{2022MNRAS.509.4886R}. We would like to note that instead of analytically modelling the likelihood function, one can also impose a synthetic likelihood. This can be done for example using a density estimate constructed using a Self-Organizing Map \citep[see, e.g.,][]{kohonen-self-organized-formation-1982} that is trained on calibration data as in e.g.\ \citet{2019arXiv191007127A, 2019MNRAS.483.2801S, 2021MNRAS.505.4249M}. In this case the same considerations would apply, where we can substitute the analytical likelihood with a likelihood that is empirically estimated. One of the methods considered, but ultimately not selected in this work is the Mizuki SED fitting method \citep{2015ApJ...801...20T, 2018PASJ...70S...9T}. Mizuki is an SED fitting technique that formulates an analytic likelihood function, so the techniques described in this section directly apply. In Appendix~\ref{sec:derivation_inference_scheme} we provide a detailed description of our sample redshift inference methodology. 

\subsection{Conditional density estimation approach}
\label{subsec:conddens}
%\rachel{Need at least one reference in this section.}\markus{done}
The conditional density estimation approach \citep[see e.g.][]{2008MNRAS.390..118L, 2013MNRAS.432.1483C, 2015MNRAS.452.3710R, 2020A&C....3000362D} constructs a density estimate between the photometry of galaxies and the redshift $p(z | \mathbf{f})$ using a calibration, or training, dataset. As such, the conditional density estimation approach depends on the calibration dataset to constrain the conditional distribution $p(z | \mathbf{f})$. The calibration dataset provides information about the mapping between photometry and redshift and the probability density of redshift given photometry.

In contrast, forward modelling explicitly considers a likelihood function or, alternatively, constructs a sampling distribution using numerical simulations. The forward modelling approach therefore must include information on the relative abundance of galaxies of different type and redshift into the prior (or as part of the simulation draws). Imposing a prior on the population distributions such as the $p_{\rm samp}(z)$ effectively acts as a regularization\footnote{Note that the likelihood is not a probability density, but a function. The probability measure is `provided' by the prior.}.

For the conditional density estimation approach, one can formulate an estimate for the sample redshift distribution via marginalization
\begin{equation}
    p_{\rm samp}(z) = \int d\mathbf{f} \, p(z | \mathbf{f}) p(\mathbf{f}) \, .
    \label{eq:marg_direct_mapping}
\end{equation}
Eq.~\eqref{eq:marg_direct_mapping} also describes a linear system, similar to Eq.~\eqref{eq:simple_linear}. However Eq.~\eqref{eq:marg_direct_mapping} is typically much better `conditioned' than Eq.~\eqref{eq:simple_linear}, if we do not consider regularization.

However due to the dependency of a conditional density estimate on a training dataset, the conditional density estimation approach often suffers from non-negligible epistemic (i.e., model) uncertainty and bias in the construction of the conditional density estimates $p(z | \mathbf{f})$. This can lead to sub-optimal probability calibration of the estimates $p(z | \mathbf{f})$. Appendix~\ref{sec:methodology_direct_mapping} describes an estimating function approach that allows the marginalization over the epistemic (or `model uncertainty') and aleatoric (or `intrinsic statistical noise') uncertainty in the estimator construction of Eq.~\eqref{eq:marg_direct_mapping}. This is achieved via the formulation of a likelihood function.

\section{Photometric Redshift Inference Pipeline}
\label{sec:photometric_redshift_inference_pipeline}
In the following subsections we describe in more detail our methodology for performing $p_{\rm samp}(z)$ inference for HSC Y3 Weak Lensing analyses.
We reiterate that all estimates for $p_{\rm samp}(z)$ in this work include the lensing weights that are available for all galaxies in the shear catalogue as described in \S~\ref{subsec:hsc_data}.

\subsection{Individual Galaxy Redshift Estimation}
\label{subsection:indiv_galax_z}
In the following, we will briefly describe the three photometric redshift techniques for individual galaxies used in this work. For a more detailed description of these methods we refer to the photometric redshift analysis study for the third public data release\footnote{\url{https://hsc-release.mtk.nao.ac.jp/doc/wp-content/uploads/2022/08/pdr3_photoz.pdf} (Accessed Oct. 2022)}. 
\subsubsection{Mizuki}
The photometric redshift code Mizuki  \citep{2015ApJ...801...20T, 2018PASJ...70S...9T} is a Spectral Energy Distribution fitting technique. It uses an SED template set constructed using Bruzual-Charlot models \citep{2003MNRAS.344.1000B}, a stellar population synthesis code that uses an Initial Mass Function following \citet{2003PASP..115..763C}, a dust attenuation modelling from \citet{2000ApJ...533..682C} and emission line modeling assuming solar metallicity \citep{10.1111/j.1365-2966.2011.18906.x}. The method applies a set of  redshift-dependent Bayesian priors on the physical properties. After estimation, the photometric redshift distributions of galaxies are calibrated \citep{2010MNRAS.406..881B} using the specXphot dataset to improve error quantification. We refer the reader to \citet{2015ApJ...801...20T} and \citet{2018PASJ...70S...9T} for more details on the methodology.
\subsubsection{DNNz}
DNNz is a neural network-based photometric redshift conditional density estimation code. The DNNz architecture consists of multi-layer perceptrons with 5 hidden layers. The training uses cmodel fluxes, unblended convolved fluxes, PSF fluxes and galaxy shape information. The construction of the conditional density uses 100 nodes in the output layer, that each represent a redshift histogram bin spanning from z = 0 to 7 (Nishizawa et al. in prep.).

\subsubsection{DEMPz}
The Direct Empirical Photometric redshift code (DEMPz) is an empirical technique for photometric redshift estimation \citep{2014ApJ...792..102H, 2018PASJ...70S...9T} that constructs conditional density estimates. The technique uses quadratic polynomial interpolation of 40 nearest neighbor galaxies in a training set, with a distance estimated in a 10 dimensional feature space (5 magnitudes, 4 colors, and shape information). DEMPz obtains error estimates for the constructed conditional densities using resampling procedures. This also includes resampling of the input feature uncertainties and bootstrapping the training galaxies.

\subsection{Sample Selection}
\label{subsec:sample_selection}
We bin the full sample described in \S~\ref{subsec:hsc_data} into four
tomographic bins by selecting galaxies using the best estimation of the DNNz
conditional density estimates within redshift intervals of (0.3, 0.6], (0.6,
0.9], (0.9, 1.2] and (1.2, 1.5].

After catalog creation we identify regions of data space that will be difficult
to calibrate using the cross-correlations with the CAMIRA LRG sample, and
therefore have the potential to produce a large systematic error (see
\S~\ref{subsec:joint_inference_camira}). In particular, we identify double solutions in the
Mizuki SED fits and DNNz conditional density estimates, associated with a significant fraction of outliers at $z\gtrsim3.0$\, for both methods.
These photometric redshift solutions have redshift-template degeneracies that produce multiple solutions. Since the secondary solutions are outside
the redshift coverage of the CAMIRA LRG sample, they cannot be calibrated using spatial cross-correlations.
Therefore, we decide to remove these galaxies from the sample.

We identify galaxies with double solutions by defining the following selection metric based on the distance between the $0.025$ and $0.975$ quantiles of the Mizuki posterior
solutions and DNNz conditional density estimates:
\begin{equation}
    \left(z_{\text{0.975}, i}^\text{Mizuki} - z_{\text{0.025}, i}^\text{Mizuki}\right) < 2.7
    \quad \text{and} \quad
    \left(z_{\text{0.975}, i}^\text{DNNz} - z_{\text{0.025}, i}^\text{DNNz}\right) < 2.7 \,,
    \label{eq:selection}
\end{equation}
where $z_{\text{0.975}, i}^\text{Mizuki}$ and $z_{\text{0.025}, i}^\text{Mizuki}$
denote the $0.975$ and $0.025$ percentiles for galaxy $i$ derived using the Mizuki
estimates of posterior redshift, respectively; and similarly for the DNNz conditional density
redshift predictions. We found that the above criteria based on the Mizuki and DNNz methods is optimal to ensure that the removal of double solutions is efficient for Mizuki, DNNz and DEMPz.

\begin{figure*}
    \centering
    \includegraphics[width=0.95\textwidth]{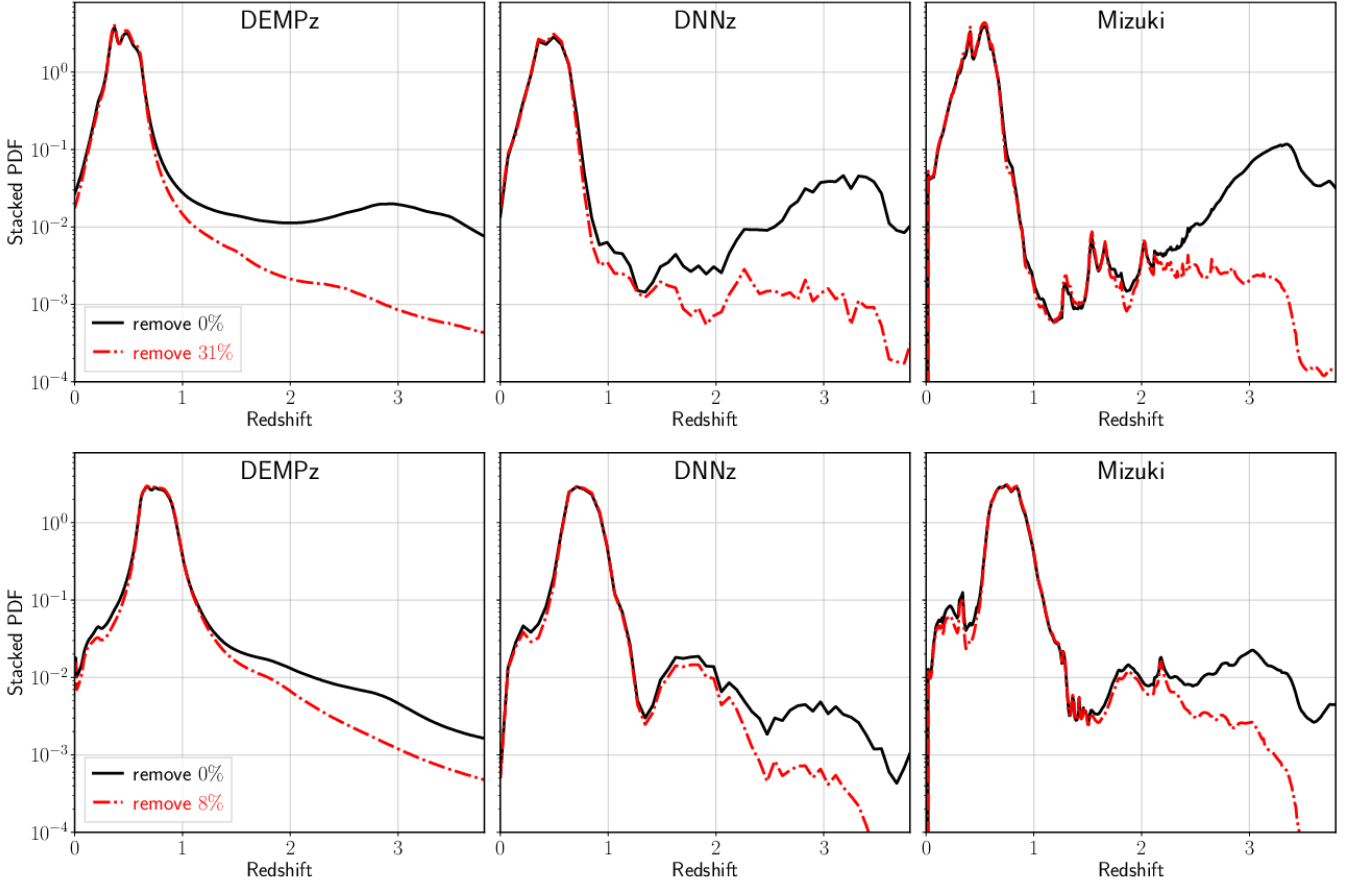}
    \caption{
    The stacked photo-$z$ posteriors for galaxies in the first (upper panel, $0.3<z_{\text{dnnz$\_$best}}\leq0.6$) and second (lower panel, $0.6<z_{\text{dnnz$\_$best}}\leq0.9$) tomographic redshift bin estimated from three photo-$z$
    estimation codes. Cuts on the interquartile distance are applied to
    these galaxies to remove the secondary peak in the stacked posteriors. The
    stacked posteriors for the fiducial cut, which removes $31\%$ of the
    galaxies in the first bin, are plotted as red lines. These posteriors are normalized so that they have total probability of
    one. 
    }
    \label{fig:stack_pzbin1_doublep}
\end{figure*}
We apply this criterion to the first and the second tomographic redshift bins,
reducing their sample size by $31\%$ and $8\%$, respectively. The third and fo\referee{u}rth tomographic bins have negligible double solutions. \referee{We, therefore, do not apply any cuts to the corresponding galaxy samples}. We illustrate the effect of removing the double solutions on the stacked (summed) redshift distribution in Fig.~\ref{fig:stack_pzbin1_doublep}. We can see that a reduction of 31\% in sample size by applying Eq.~\eqref{eq:selection} removes double solutions for all three methods available in this work. In the following, we will
denote the removal of double solutions as the `calibration cut'.

We have also confirmed that this selection does not induce a spatial selection effect. This was tested by comparing the spatial distribution of galaxies before and after we apply the calibration cut and confirming that no significant modification of the clustering was introduced by the cut.

\begin{figure*}
    \centering
    \includegraphics[scale=0.5]{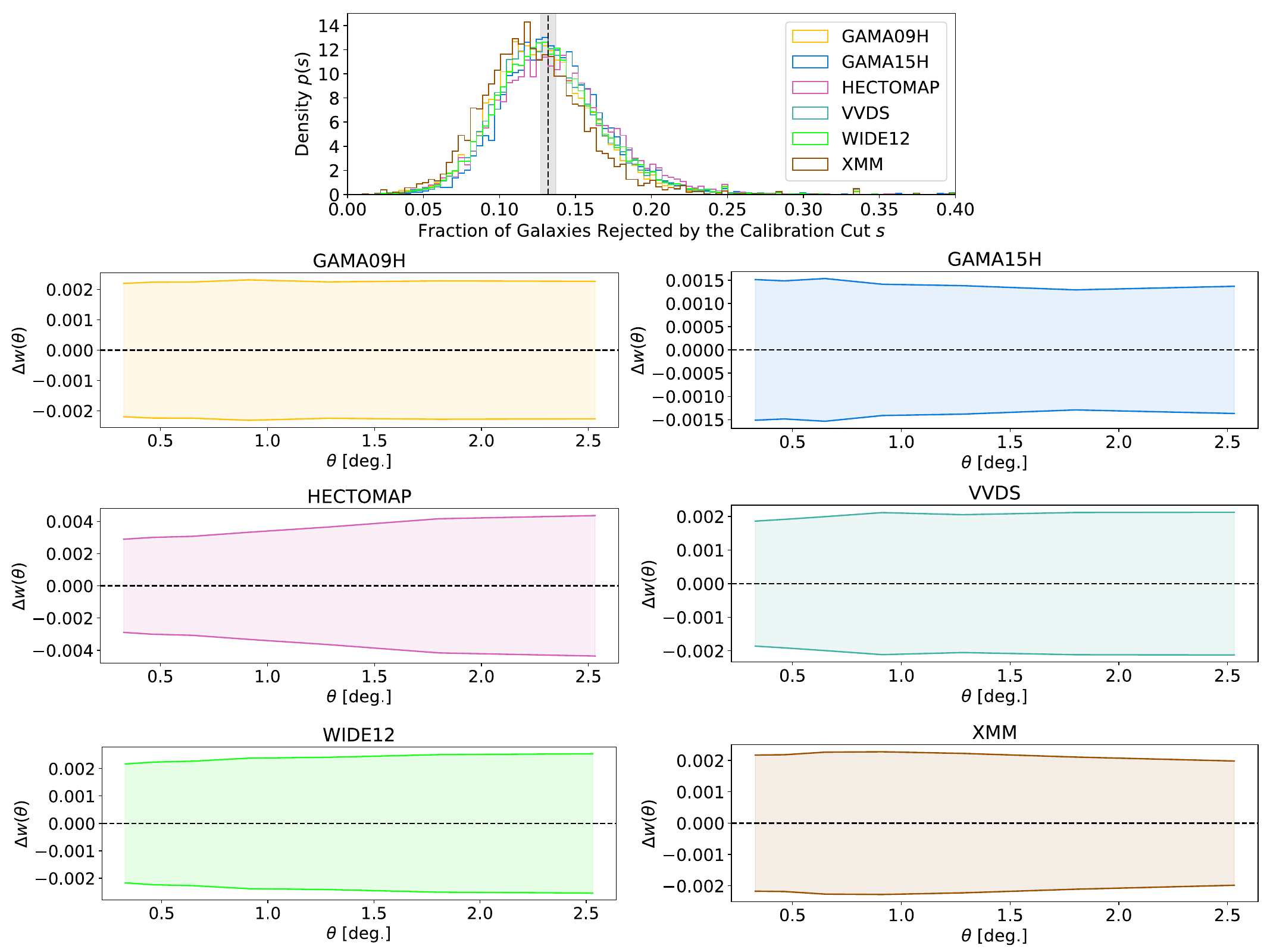}
    \caption{Testing the impact of the calibration cut on the spatial distribution of galaxies by resampling the catalog for the first tomographic bin. \textit{Top Panel: } Test if the fraction of galaxies rejected by the calibration cut ($s$) is comparable for all subfields. Each histogram corresponds to a separate field listed in the legend, where the histograms over $s$ show the variation across the different patches within the field. The vertical dashed line denotes the mean of the histograms over the different fields with errorbars denoting the field-to-field variation. \textit{Lower Panels: } Testing if the spatial distribution of removed galaxies is `random'. The vertical axis shows the difference between the correlation function estimated on the catalog in each field subject to the calibration cut and a catalog where galaxies are removed randomly. The horizontal dashed line shows the zero line. The error contours are obtained by jackknife resampling.   
    }
    \label{fig:spatial_var}
\end{figure*}

This is illustrated in Fig.~\ref{fig:spatial_var}, where we test the impact of the calibration cut on the spatial distribution of galaxies. We first confirm if the fraction of galaxies rejected by the calibration cut (i.e., galaxies with doubly-peaked $p_{\rm indiv}(z)$) $s$ is comparable for all subfields. This has to take into account the variation due to sampling variance, which we quantify by dividing into subregions 
%\rachel{should this be changed to `by dividing into subregions'} 
within the different fields. The top panel plots several normalized histograms over $s$ where each histogram corresponds to a separate field listed in the legend. Note that we obtain a distribution $p(s)$ over $s$ for each field by estimating $s$ on each patch within each field. The vertical dashed line denotes the mean of the histograms over the different fields, the errorbars denote the field-to-field variation. We see that $s$ is consistent across the different fields.

In the lower panels we investigate if the spatial distribution of removed galaxies is spatially `random', or if we have to expect a correlation signal based on the calibration cut. The vertical axis shows the difference between the correlation function estimated on the catalog in each field subject to the calibration cut and a catalog where galaxies are removed randomly. The horizontal dashed line guides the eye towards the zero line. The error contours are obtained by jackknife resampling the catalog within each field. We see that the measured autocorrelation functions are consistent between the randomly selected catalog and the catalog subject to the calibration cut.   
% \rachel{I find this confusing: is this a statement that you have divided the regions up into $N_p$ patches and the histogram values come from estimating $s$ after successively excluding each patch?  (i.e., traditional jackknife)  If so: I'm confused how $s$ can vary so much when the samples are highly correlated.} 
%     \textcolor{red}{(AJN reply: You are right but the \textit{s} is estimated from each patch,i.e, single estimation of $s$ is from single patch instead of the jackknife subsample.)}
\subsection{Individual Galaxy Redshift Estimation to enable Sample Redshift Distribution ($p_{\rm samp}(z)$) Inference}
\label{indiv_galaxy_pz_estimation}
This project considered all three individual galaxy photometric redshift estimates introduced in \S~\ref{subsection:indiv_galax_z} and performed an initial comparison between sample redshift posteriors obtained using these three methods with the cross-correlation constraints. We found insufficient agreement for the Mizuki solutions, whereas DEMPz and DNNz where more consistent. By iteratively reproducing the inconsistencies using analytic error models, we identified a number of problems with the Mizuki solutions. 

We found that the Mizuki photometric redshift solutions are miscalibrated (Nishizawa et al. in prep.) and that systematics induced by uncorrected selection functions from galaxy selection, object weighting and the calibration cut can lead to additional bias in the sample redshift inference for the Mizuki code. A recalibration of the Mizuki likelihoods using the specXphot sample based on \citet{2010MNRAS.406..881B} only slightly improved the results. We concluded, that the consistency between the DEMPz and DNNz codes and the cross-correlation measurements was still better. We note that including the aforementioned selection function into the likelihood formulation is structurally simple, but would require a rerun of the Mizuki solutions which was not deemed practical. We therefore selected DNNz as our primary method and DEMPz as the alternative method for the subsequent analysis. In the following, we will refer to sample redshift distribution inference methodology based on individual galaxy redshift distributions, abbreviated as the vector-valued $\vec{p}_{\rm indiv}(z)$, as `photometry-based $p_{\rm samp}(z)$ estimation', or short `PhotZ'.

\subsection{Formulation of the Ensemble Redshift Distribution Prior}
\label{subsec:sample_pz_PhotZ}

Based on our fiducial model choice we apply the empirical likelihood methodology described in Appendix~\ref{sec:methodology_direct_mapping} to estimate $p_{\rm samp}(z)$ for the four tomographic bins based on the DNNz $\vec{p}_{\rm indiv}(z)$.

As we discuss in detail in Appendix~\ref{sec:methodology_direct_mapping}, the empirical likelihood estimation obeys the central limit theorem. The large sample size of our catalogs implies that the statistical error in the maximum empirical likelihood estimate is much smaller than other sources of uncertainty. These include a cosmic variance contribution from the spatially limited training sample (see \S~\ref{subsec:spec_reference}), as well as the uncertainty in the individual galaxy redshift estimation model (epistemic uncertainty). In the remainder of this section we will discuss our approach to including cosmic variance into our sample redshift estimation procedure. Our treatment of the epistemic uncertainty will be discussed in \S~\ref{subsec:prior_recommendation_wl}.

The basis for our $p_{\rm samp}(z)$ error model is the logistic Gaussian process. The logistic Gaussian process, first applied to sample redshift estimation by \citet{2020MNRAS.491.4768R}, assumes that the number counts of galaxies as a function of redshift are lognormally distributed. The model can capture cross-bin correlations and provides more modelling complexity than, e.g., the Dirichlet distribution as we discuss in Appendix~\ref{sec:PriorChoice}.

The logistic Gaussian process prior on the parameters $\boldsymbol{\phi_\text{nz}}$ can be formulated as follows:
\begin{equation}
\begin{split}
\boldsymbol{s} &\sim \mathcal{N}(\boldsymbol{s} | \mu, \Sigma) \\
\boldsymbol{\rho} &= \exp(\boldsymbol{s}) \\
    \boldsymbol{\phi_\text{nz}} &:= \left\{\frac{\rho_i}{\sum_j \rho_j}  \ \bigg|  \ 0 < i < N_{\rm bins} \right\}   \, ,
    \end{split}
    \label{eq:logGPsimple}
\end{equation}
where ($\boldsymbol{\mu}$/$\boldsymbol{\Sigma}$) denotes the (mean vector/covariance matrix). We note
that Eq.~\eqref{eq:logGPsimple} relates to a lognormal model for the galaxy counts, where $\rho$ is the expected number of galaxies per redshift. The dimension of ($\boldsymbol{s}$/$\boldsymbol{\rho}$/$\boldsymbol{\phi_\text{nz}}$) is $N_{\rm bins}$ as introduced in Eq.~\eqref{eq:parametrization_pz}. 

As discussed in \S~\ref{subsec:spec_reference}, the faint end of our training set is dominated by COSMOS2015 data. This induces a cosmic variance error contribution that we include into our logistic Gaussian process model based on the cosmic variance measurements for the COSMOS2015 dataset by \citet{2020arXiv200409542S}. We detail our methodology in Appendix~\ref{sec:characterizing_cv_loggp}.

\subsection{Ensemble Redshift Distribution Likelihood from Spatial Cross-Correlations (Cross-Correlation)}
\label{subsec:joint_inference_camira}
To further constrain the $p_{\rm samp}(z)$, we utilize spatial cross-correlations with the CAMIRA LRG sample. This approach has two goals: it provides an independent consistency check for the $p_{\rm samp}(z)$ derived using the DNNz approach, and it allows a joint inference of the $p_{\rm samp}(z)$ informed by both the photometry of galaxies and the spatial cross-correlations with the CAMIRA LRG sample.

As detailed in \S~\ref{subsec:camira_data}, the CAMIRA LRG sample extends only to $z \lesssim 1.2$ and the photoZ of the CAMIRA LRG galaxies are themselves subject to error. This subsection gives an overview of the cross-correlation measurements and the likelihood formulation. We refer to Appendix~\ref{sec:marg_camira_lrg} for the technical details.

Using vector notation, where each vector component corresponds to the cross-correlation measurement in a redshift bin, we can predict the spatial cross-correlation between the CAMIRA LRG sample and HSC phot as
\begin{equation}
    \mathbf{w_{\rm LRG-Y3}} \propto \boldsymbol{\phi_\text{nz}} \, \mathbf{b_{\rm PhotZ}} \,  \mathbf{b_{\rm LRG}}  \, \mathbf{w_{\rm DM}} \, ,
    \label{eq:wx_definition}
\end{equation}
where $\mathbf{w_{\rm DM}}$ is the scale averaged, redshift and cosmology dependent, two-point function of the dark matter density field. The terms $\mathbf{b_{\rm PhotZ}}$ and $\mathbf{b_{\rm LRG}}$ are the redshift-dependent galaxy-dark matter bias terms from the (HSC phot/CAMIRA LRG) sample and $\boldsymbol{\phi_\text{nz}}$ are the parameters defined in Eq.~\eqref{eq:parametrization_pz}. We use `The-Wizz' \citep[a code described in][]{Morrison2016} to measure these cross-correlations and use bootstrap re-sampling (as described in \citealt{Morrison2016}) to obtain a covariance matrix of the measurements. We include the lensing weights in the two-point estimator, and choose a scale range of $0.1-1.0 \, {\rm Mpc}$ for our measurements. These measurements are repeated for 10 catalogs generated by sampling from our CAMIRA LRG photometric error model, which is a conditional density estimate that maps the noisy CAMIRA LRG photometric redshift to the unknown true redshifts. This mapping is trained on the specXphot calibration data.

Using the scheme described in Appendix~\ref{sec:marg_camira_lrg} we marginalize over the realisations to derive a likelihood for the cross-correlation measurements that has an inflated covariance $\boldsymbol{\Sigma}_{\rm LRG-PhotZ}$ due to the contribution of the CAMIRA LRG photometric redshift error. Using a Gaussian Likelihood ansatz we obtain
\begin{equation}
\begin{split}
    &p(\mathbf{\hat{w}_{\rm LRG-PhotZ}} | \boldsymbol{\phi_\text{nz}}_{\rm PhotZ}, \mathbf{b_{\rm PhotZ}}, \mathbf{b_{\rm LRG}})  \\&= \mathcal{N}(\mathbf{\hat{w}_{\rm LRG-PhotZ}} | \mathbf{w}_{\rm LRG-PhotZ}(\boldsymbol{\phi_\text{nz}}_{\rm PhotZ}, \mathbf{b_{\rm PhotZ}}, \mathbf{b_{\rm LRG}}), \boldsymbol{\Sigma}_{\rm LRG-PhotZ})
    \end{split}
    \label{eq:wx_like}
\end{equation}
where $\mathbf{\hat{w}_{\rm LRG-PhotZ}}$ denotes the spatial cross-correlation measurements between the CAMIRA LRG and HSC phot catalogs, $\mathbf{w}_{\rm LRG-PhotZ}(\boldsymbol{\phi_\text{nz}}_{\rm PhotZ}, \mathbf{b_{\rm PhotZ}}, \mathbf{b_{\rm LRG}})$ denotes the theory prediction and $\boldsymbol{\Sigma}_{\rm LRG-PhotZ}$ the covariance matrix that is adjusted for the CAMIRA LRG photometric redshift error.

In this analysis we marginalize over a parameter that describes the product $\mathbf{b_{\rm PhotZ}} \, \mathbf{b_{\rm LRG}}$ for each tomographic bin. For three tomographic bins we therefore have three parameters that account for the product of galaxy-dark matter bias for galaxies in the HSC phot and the CAMIRA LRG samples. We predict\footnote{We use: $\Omega_{\rm DM} = 0.258868$, $\Omega_{\rm b} = 0.048252$, $h = 0.6777$, $n_s = 0.95$ and $\sigma_8 = 0.8$.} the dark matter contribution $\mathbf{w_{\rm DM}}$ using the \textit{Core Cosmology Library, version 1.0.0} (CCL, \citealt*{2019ApJS..242....2C})\footnote{\url{https://github.com/LSSTDESC/CCL} (Accessed: 09/22/2022)} using halofit to model the nonlinear power spectrum \citep{2012ApJ...761..152T}.  
We do not marginalize over cosmological parameters that enter $\mathbf{w}_{\rm LRG-PhotZ}$, as we find that the choice of cosmology does not strongly impact the posterior $p_{\rm samp}(z)$. Concretely, we note that the spatial cross-correlation data vector is a scale-averaged correlation function. Its redshift scaling affects the inferred cross-correlation redshift distributions on the $\sim 20\%$ level by (suppressing/increasing) the (low/high)-z flank. However variations in cosmology affect the redshift scaling of the scale-averaged dark-matter correlation at the $\sim 10\%$ level (for rather extreme cosmologies at the $2\sigma$ contour of Stage III surveys), which implies that the cosmology-dependence of the inferred cross-correlation redshift distributions is subdominant to other systematics such as the redshift-dependent galaxy-dark matter bias modelling uncertainties. 

\subsection{Joint Constraints}
\label{subsec:joint_constraint}
\begin{figure*}
    \centering
    \includegraphics[scale=1.2]{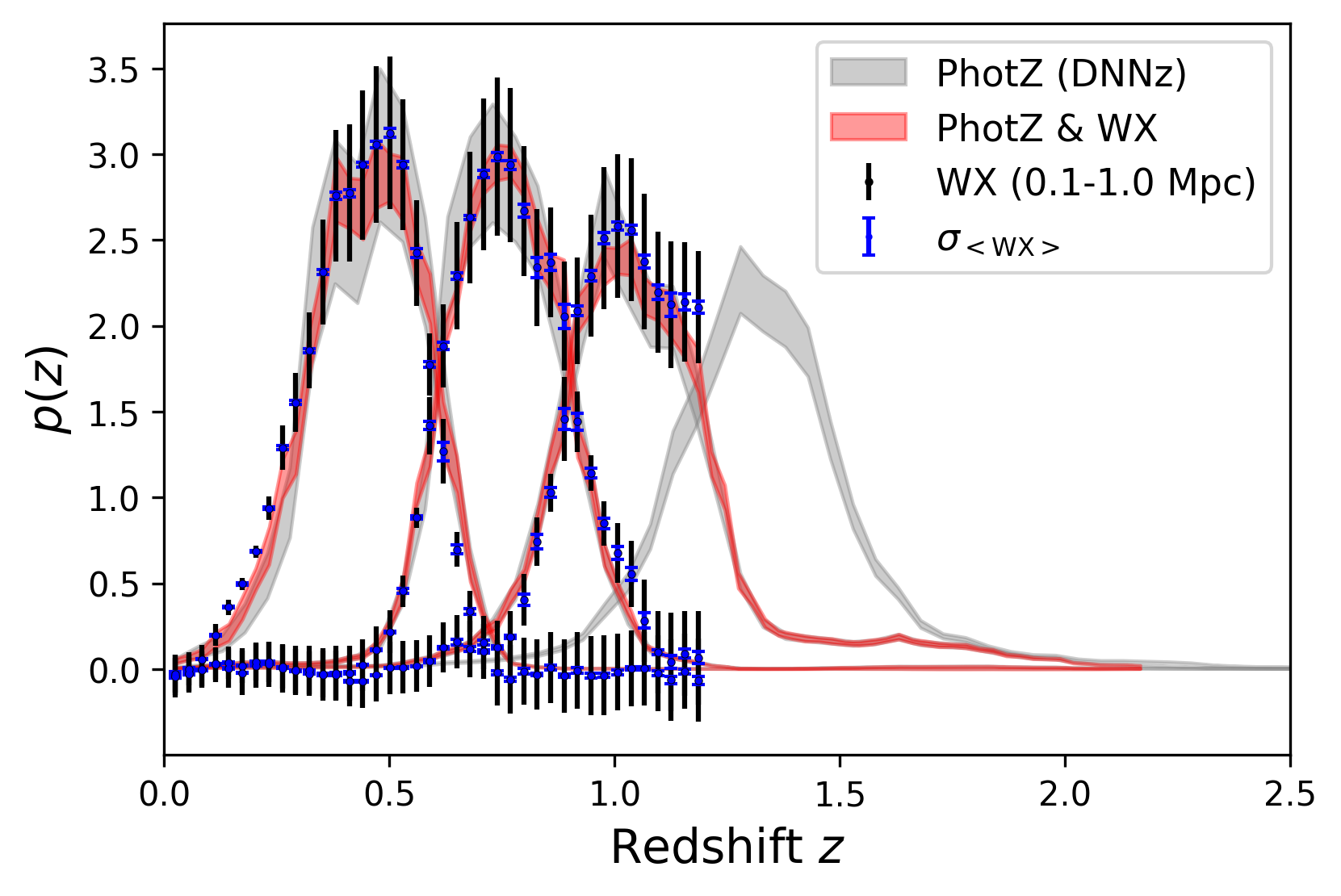}
    \caption{Sample redshift distribution ($p_{\rm samp}(z)$) posteriors for the four tomographic
    redshift bins of the HSC Y3 lensing sample derived to include
information from the photometry (`PhotZ (DNNz)', grey area), spatial clustering (`WX (0.1 - 1.0 {\rm Mpc})',
black dots) and the combination of spatial information and photometry (`PhotZ \& WX', red area). The blue dots denotes the standard deviation on the mean of WX (i.e., clustering redshift) measurements. 
The CAMIRA LRG sample has a limited redshift coverage to $z < 1.2$, due to which the high redshift tomographic bin does not include a cross-correlation data vector. The inference includes the lensing weights consistently in all likelihood terms. The piecewise intervals denote the $\pm 1 \sigma$ errors. }
    \label{fig:joint_inference}
\end{figure*}
Using the logistic Gaussian Process model defined in \S~\ref{subsec:sample_pz_PhotZ} and the cross-correlation likelihood defined in Eq.~\eqref{eq:wx_like}, we can sample from the joint posterior of the parameters that describe the sample redshift distribution $\boldsymbol{\phi_\text{nz}}$, defined in Eq.~\eqref{eq:parametrization_pz}, and the product $\boldsymbol{b} = \boldsymbol{b}_{\rm LRG} \, \boldsymbol{b}_{\rm PhotZ}$ of the galaxy-dark matter bias of the CAMIRA LRG ($\boldsymbol{b}_{\rm LRG}$) and HSC phot ($\boldsymbol{b}_{\rm PhotZ}$) samples
\begin{equation}
    p(\boldsymbol{\phi_\text{nz}}, \boldsymbol{b} | \mathbf{\hat{w}_{\rm LRG-PhotZ}} )
     \propto p(\mathbf{\hat{w}_{\rm LRG-PhotZ}} | \boldsymbol{\phi_\text{nz}}, \boldsymbol{b}) p(\boldsymbol{\phi_\text{nz}})p( \boldsymbol{b}) \, .
    \label{eq:joint_like}
\end{equation}
The sampling of the $\boldsymbol{\phi_\text{nz}}$ parameters has to be carried out with respect to a likelihood that only constrains a subset of $\boldsymbol{\phi_\text{nz}}$ due to the limited redshift coverage of the CAMIRA LRG sample. We note that the parameters $\boldsymbol{\phi_\text{nz}}$ can be normalized to lie on the simplex\footnote{The probability simplex is defined as $\mathcal{S} = \{x_i | \sum_{i = 1}^{N} x_i = 1 \mbox{~and~} 0 \leq x_i \leq 1 \mbox{~for~} 1 \leq i \leq N \}$.}, i.e., to sum to unity. It is, therefore, useful to instead perform inference with respect to the random variable $\mathbf{s}$, defined in Eq.~\eqref{eq:logGPsimple}. Using this reparametrization we can perform inference in $\mathbb{R}^{\rm N_{\rm bins}}$ using standard approaches and then transform to the original parameter $\boldsymbol{\phi_\text{nz}}$. We use Elliptical Slice Sampling \citep{pmlr-v9-murray10a} for our inference. Elliptical slice sampling works particularly well for a logistic Gaussian process prior, since it can utilize the aforementioned reparametrization that relates the logistic Gaussian process to the multivariate normal distribution.

Fig.~\ref{fig:joint_inference} shows the resulting posterior sample redshift
distributions for the following three scenarios:
\begin{itemize}
    \item photometry-based sample redshift distribution estimation (`PhotZ (DNNz)', grey) utilizing the DNNz code and including our model for cosmic variance following \S~\ref{indiv_galaxy_pz_estimation} and \S~\ref{subsec:sample_pz_PhotZ};
    \item clustering redshift estimation (`WX (0.1 - 1.0 Mpc)', black) following \S~\ref{subsec:joint_inference_camira}; and
    \item  the combination of spatial information and photometry (`PhotZ \& WX', red) following \S~\ref{subsec:joint_constraint}.
\end{itemize}
The horizontal axis of Fig.~\ref{fig:joint_inference} shows the redshift, while the vertical axis shows the probability density of posterior tomographic $p_{\rm samp}(z)$. The distributions are normalized to integrate to unity. We report contours/errorbars corresponding to piecewise $\pm 1 \sigma$ errors. In the case of `PhotZ' and `PhotZ \& WX' which both have asymmetric posterior distributions, we report contours between the 16th and 84th percentiles.
The blue errorbars show the standard deviation in the mean\footnote{We refer here to the standard deviation in the mean estimate, that scales with $1/\sqrt{N}$, where N corresponds to the number of catalogs drawn.} cross-correlation measurement with respect to the different catalog draws from the CAMIRA LRG error model. We specifically see that even for only 10 catalogs, this error is already much smaller compared with the statistical uncertainty of `WX (0.1 - 1.0 Mpc)'. We note that the black errorbars for the cross-correlation constraints are plotted assuming the maximum a-posteriori values of $\boldsymbol{b}$ defined in Eq.~\eqref{eq:joint_like}, which act to normalize the clustering redshift measurements. This allows us to plot the clustering redshift constraints on the same scale as `PhotZ (DNNz)' and `PhotZ \& WX'. We note that we do this for illustrative purposes only; we marginalize over $\boldsymbol{b}$ to infer `PhotZ \& WX'.

Since the CAMIRA LRG sample redshift
coverage extends to $z < 1.2$, we can only partially calibrate the third tomographic bin. \referee{We also decided to not include a cross-correlation data vector in the sample redshift distribution calibration of the fourth tomographic bin. This is motivated by the overall small redshift overlap with the CAMIRA LRG sample. Furthermore, for significant parts of the relevant redshift range ($1.0 < z < 1.2$), there is a trend in the inferred $n(z)$ in the third bin that might indicate the need for more complex modelling of astrophysical effects like redshift dependent galaxy-dark matter bias. It is therefore likely that we might include additional systematics in the calibration of the fourth tomographic bin low-redshift tail for very moderate gains in statistical accuracy.   }

We conclude that the clustering redshift measurements are broadly consistent with the constraints we derive based on the photometry of galaxies. However there are slight inconsistencies between the `PhotZ' and 'WX' constraints around $z \approx 0.2$. This is around the same redshift where we know that the photometric redshift distributions of the CAMIRA LRG galaxies are biased (see \S~\ref{subsec:camira_data}). 
This implies an incomplete correction of this bias from our error model. This inconsistency is moderate, on the level of $2\sigma-3\sigma$ with respect to the joint posterior (`PhotZ \& WX'). We leave further investigations for future work. 

\subsection{Prior Recommendation for Weak Lensing Analysis}
\label{subsec:prior_recommendation_wl}
\begin{figure*}
    \centering
    \includegraphics[scale=1.2]{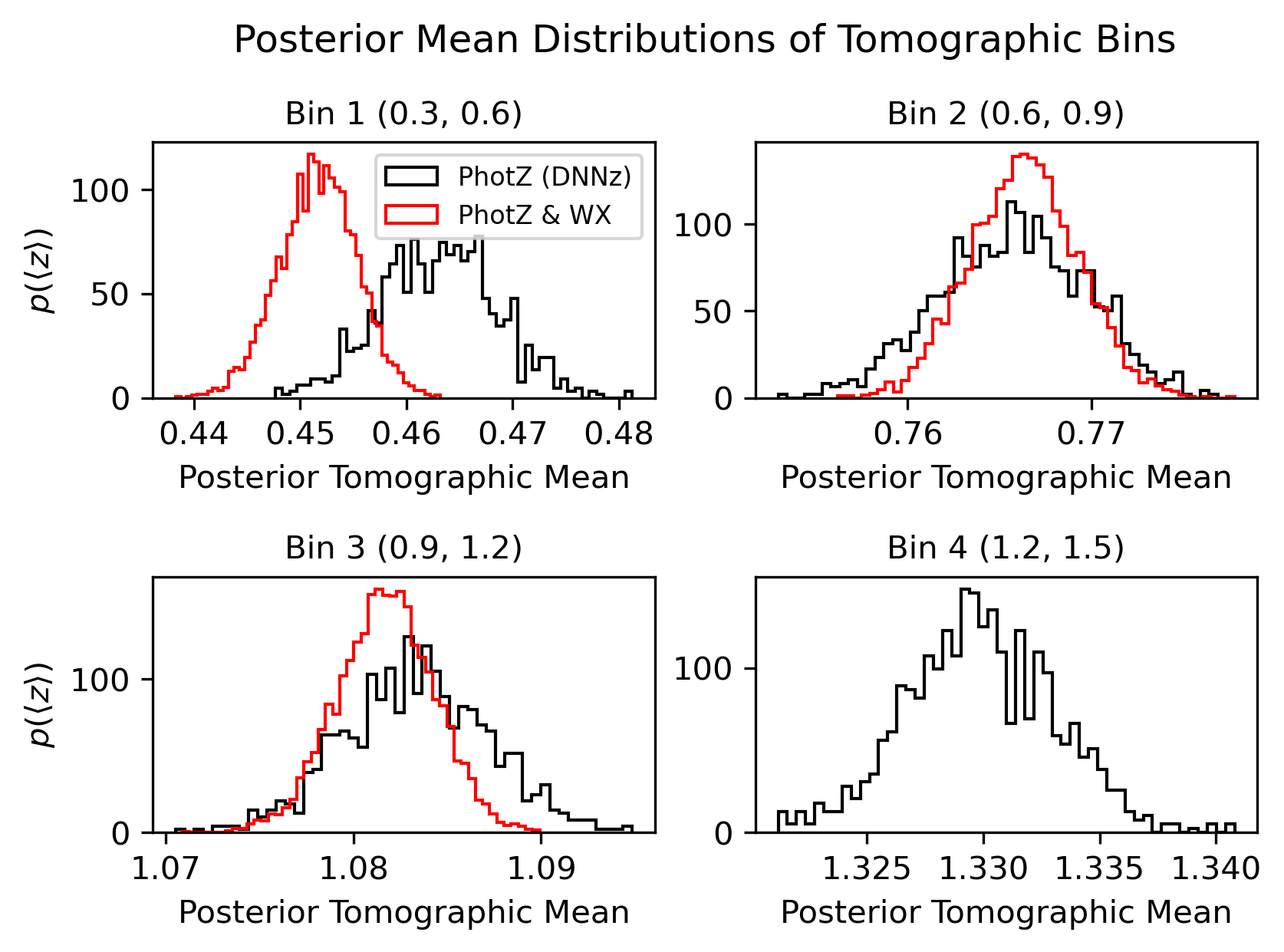}
    \caption{ Comparison of the distributions of posterior tomographic mean for
    the four tomographic redshift distributions shown in
Fig.~\ref{fig:joint_inference_post_mean}. The subpanels correspond to increasing
tomographic bin mean redshift. The (red/black) histograms
show the result for the constraint (`PhotZ (DNNz)'/`Phot \& WX') which corresponds the (exclusion/inclusion) of the spatial cross-correlation datavector with the CAMIRA LRG sample. %\markus{
There is consistency between the posterior distributions of tomographic mean estimates obtained using the photometry alone (black) and in combination with the clustering redshift datavector (red).
}
    \label{fig:joint_inference_post_mean}
\end{figure*}
\begin{figure}
    \centering
    \includegraphics[scale=0.53]{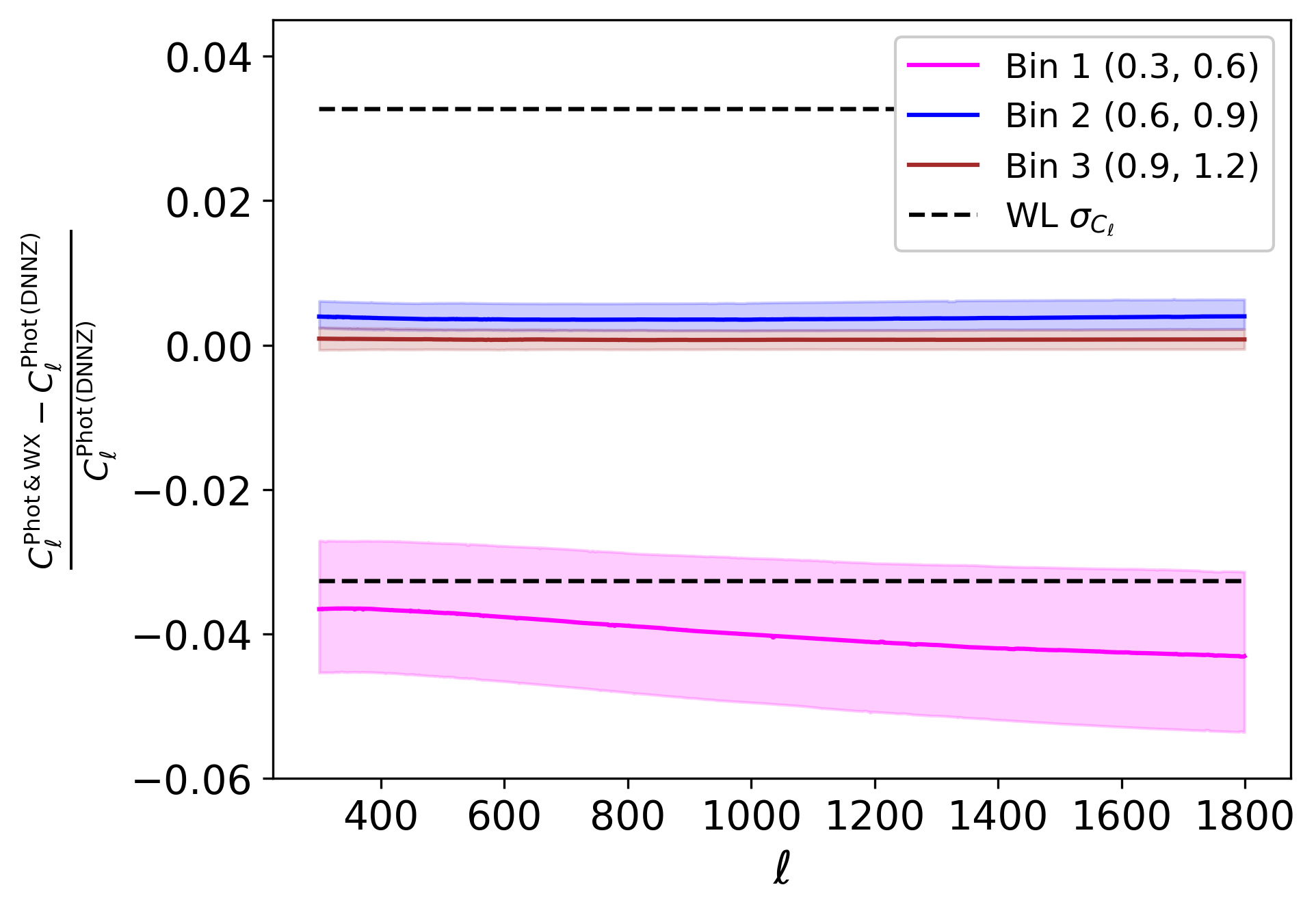}
    \caption{ Comparison of the distributions of relative bias in weak lensing power spectra (see Eq.~\ref{eq:rel_bias_cl}) between the posterior $p_{\rm samp}(z)$ informed by the photometry alone (Phot (DNNz)) and the joint constraints that include the spatial cross-correlations (Phot \& WX). We plot the results for the (first/second/third) tomographic bins (Bin 1/Bin 2/Bin 3) corresponding to the results shown in Fig.~\ref{fig:joint_inference} and Fig.~\ref{fig:joint_inference_post_mean}. The solid lines show the median and the contours show the (16/84) percentiles corresponding to the Gaussianized $\pm 1 \sigma$ errors. The black horizontal dashed lines show the $\pm 1 \sigma$ errors that correspond to the expected signal-to-noise ratio of the weak lensing power spectra measurements.   }
    \label{fig:clbias}
\end{figure}
Fig.~\ref{fig:joint_inference_post_mean} shows the distribution of posterior
mean for the four tomographic bins.
We define the posterior mean as the mean estimated for each posterior tomographic $p_{\rm samp}(z)$ sample. We can derive the distribution of posterior mean for each tomographic bin by sampling from the posterior $p_{\rm samp}(z)$ shown in
Fig.~\ref{fig:joint_inference_post_mean}. This is done for the joint constraint (`PhotZ \& WX', red contours) and the photometry-based inference (`PhotZ (DNNz)', grey contours) for each tomographic bin. We now estimate the mean of each sample drawn in this way. This results in distributions of posterior mean for our tomographic bins in both scenarios.

We see that the distributions of posterior mean are consistent for the two methods in the first three tomographic bins. There is mild tension in the lowest tomographic bin, which can be explained by the inconsistency at $z \approx 0.2$ as described in the previous section. We further quantify the `information gained' by the cross-correlation likelihood over the `PhotZ (DNNz)' prior by calculating the Kullback-Leibler (KL) Divergence between the prior and posterior based on the results quoted in Tab.~\ref{tab:summary_stats}, where we use a Gaussian approximation for the posterior mean distributions of tomographic bins to calculate the KL Divergence. The KL divergence between prior and posterior is referred to as the `Bayesian Surprise' in statistics \citep[see, e.g.,][]{ITTI20091295, BALDI2010649}\footnote{The Bayesian Surprise is sometimes referred to as the `information gain' in cosmology \citep[e.g.][]{2016JCAP...05..034G}.} and the results are quoted in Tab.~\ref{tab:summary_stats} under the column `Bayesian Surprise'. Tab.~\ref{tab:summary_stats} indicates that the largest amount of information is added in the first tomographic bin. We note, however, that this does not allow us to judge if the Bayesian Surprise is due to unaccounted systematics or statistical fluctuation. A comparison with the results from the second and third bin, which are an order of magnitude smaller, hints towards unaccounted systematics in the first bin as the most likely explanation for the large Bayesian surprise value.

Fig.~\ref{fig:joint_inference_post_mean} further illustrates that the width of the distributions of posterior mean decreases when we include the spatial cross-correlation data vector. This highlights the importance of including cross-correlations in the sample redshift calibration as both a consistency check and an additional constraint. We relate this result to the expected biases in the weak lensing power spectra in Fig.~\ref{fig:clbias}, proceeding in close analogy to our study of the distribution of posterior mean. We estimate the weak lensing power spectra on each draw from the posterior $p_{\rm samp}(z)$ using the \textit{Core Cosmology Library, version 1.0.0} %\rachel{what version?} 
(CCL, \citealt*{2019ApJS..242....2C})\footnote{\url{https://github.com/LSSTDESC/CCL} (Accessed: 09/22/2022)} and calculate the relative bias $\Delta C_{\ell}$ between the posterior distributions of weak lensing power spectra estimated using the photometry alone (Phot (DNNz)) and including the spatial cross-correlations (Phot \& WX). The relative bias is defined as
\begin{equation}
 \Delta C_{\ell} = \frac{C_{\ell}^{\rm Phot \, \& \, WX} - C_{\ell}^{\rm Phot \, (DNNz)}}{C_{\ell}^{\rm Phot \, (DNNz)}} \, .
 \label{eq:rel_bias_cl}
\end{equation}

Fig.~\ref{fig:clbias} shows $\Delta C_{\ell}$ as a function of scale for the (first/second/third) tomographic bin. We see that the relative difference between the measurements using the photometry (DNNz) alone shows a tension that is significant in Bin 1 compared with the expected signal-to-noise ratio, which hints towards remaining uncorrected systematic biases. We discuss this further in \S~\ref{sec:discussion_futurework}. In the following we discuss our conservative assessment of tomographic $p_{\rm samp}(z)$ error motivated by the aforementioned tensions.

\begin{table*}
    \centering
    \begin{tabular}{ c | c | c | c | c | c | c }
       & Y1 Analysis & Y3 PhotZ (DNNz) & Y3 DEMPz & Y3 PhotZ \& WX & Y3 Bayesian Surprise  & Y3 Total  \\\hline\hline
       Bin 1 &  0.44 (0.0285) & 0.463 (0.005) & 0.463 & 0.452 (0.004) & 3.84 &  0.452 (0.024) \\
       Bin 2 &  0.77 (0.014) & 0.766 (0.004) & 0.777  & 0.766 (0.003) & 0.10 & 0.766 (0.022) \\
       Bin 3 & 1.05 (0.0383)& 1.084 (0.004) & 1.097  & 1.081 (0.004) & 0.28 & 1.081 (0.031) \\
       Bin 4 & 1.33 (0.0376) & 1.330 (0.003) & 1.350  & - & - & 1.330 (0.034)  \\
        \end{tabular}
    \caption{Mean and standard deviation of the posterior mean for the different tomographic redshift bins. The first column lists the corresponding results for the first year analysis \citep{2020PASJ...72...16H} (`Y1 Analysis'), the results obtained using the photometry alone with cosmic variance correction (`PhotZ (DNNz)'), the results we obtain using the DEMPz code (`Y3 DEMPz') and the joint constraints with the cross-correlation data vector (`Y3 PhotZ \& WX'). The DEMPz results, here used as an alternative methodology, are obtained by taking the average of the normalized $\vec{p}_{\rm indiv}(z)$. For conditional density estimates like DEMPz this amounts to a mean estimate of the marginalization in Eq.~\ref{eq:marg_direct_mapping} (see \S~\ref{subsec:conddens}). The final two columns lists the Bayesian Surprise values (`Y3 Bayesian Surprise') and the total error budget that includes our systematics error budget as explained in \S~\ref{subsec:prior_recommendation_wl} (`Y3 Total'). We note that all columns except the first are derived on the year 3 dataset described in \S~\ref{sec:data} with different galaxy selection (but similar redshift range) compared with the S16A analysis. }
    \label{tab:summary_stats}
\end{table*}
Since the sample redshift posteriors obtained in this work will be used as part of the HSC Y3 weak lensing cosmological analysis, we discuss here which parameterization we will employ to marginalize over sample redshift uncertainty. Following \citet{2022arXiv220610169Z} we will use the maximum {\em a posteriori} solution for the $p_{\rm samp}(z)$ and vary the mean using a Gaussian prior informed by the inference described in the previous sections. While \citet{2022arXiv220610169Z} explored multiple ways of marginalizing over the full posterior for the redshift distribution, at the level of precision of this HSC analysis, marginalizing over uncertainty in the mean redshift was found to be entirely sufficient. 
We also include an additional error contribution that parametrizes differences in sample redshift inference across different $\vec{p}_{\rm indiv}(z)$ solutions, where we will use DEMPz as an alternative method.

We derive the combined error budget based on the aforementioned parametrization of the posterior mean. In order to include discrepancies between different $\vec{p}_{\rm indiv}(z)$ solutions into the analysis, we compare the results obtained using DNNz with the DEMPz results. The DEMPz method was selected because it showed superior photometric redshift accuracy compared with the Mizuki results\footnote{See \url{https://hsc-release.mtk.nao.ac.jp/doc/wp-content/uploads/2022/08/pdr3_photoz.pdf}} and overall better consistency with the clustering redshift measurements.

Since the DEMPz and DNNz methods will be correlated, we have to formulate an upper limit on the error budget. Furthermore we require that this upper limit calculation will be conservative with respect to the residual systematics in Bin 1 discussed in Fig.~\ref{fig:joint_inference_post_mean} and the  HSC first-year (Y1) result \citep{2020PASJ...72...16H} for Bin~4, as Bin~4 lacks the additional constraints from the spatial correlations with the CAMIRA LRG sample. 

While we present a significantly updated methodology, we do not provide additional data driven consistency checks that would warrant a significantly smaller systematic error budget compared with the Y1 analysis.
To derive this total error budget we combine the standard deviation of the posteriors of the joint constraint (shown as red histograms in Fig.~\ref{fig:joint_inference_post_mean}), which we will denote as $\sigma_{\rm joint}$, with the absolute difference between the $p_{\rm samp}(z)$ derived using the alternative method DEMPz and our joint fiducial analysis. The latter error contribution will be denoted as $\sigma_{\rm sys}$. We reiterate that we consider here only the posterior tomographic mean.

We introduce the correlation coefficient $\rho$ with $|\rho| \leq 1$ and combine $\sigma_{\rm sys}$ with the statistical error budget $\sigma_{\rm joint}$ as
\begin{equation}
\begin{split}
\sigma_{\rm joint, sys} &= \sqrt{\sigma_{\rm joint}^2 + \sigma_{\rm sys}^2 + 2 \rho \sigma_{\rm sys} \sigma_{\rm joint}} \\
&\leq \sqrt{\sigma_{\rm joint}^2 + \sigma_{\rm sys}^2 + 2 \sigma_{\rm sys} \sigma_{\rm joint}} \, .
\end{split}
\label{eq:combination_sys}
\end{equation}
An upper limit on $\sigma_{\rm joint, sys}$ is therefore given as $\sigma_{\rm joint, sys} \leq  \sigma_{\rm joint} + \sigma_{\rm sys}$. This implies an upper limit for (Bin 1/Bin 4) of $\sigma_{\rm joint, sys, (Bin 1/Bin 4)} = (0.015/0.023)$. This systematic error budget for the Bin 1 and Bin 4 is similar to the absolute difference between the constraints of `PhotZ (DNNz)' and `Phot \& WX' in Fig.~\ref{fig:joint_inference_post_mean} and much smaller than the error budget for Bin 4 assumed in Y1 as quoted in Tab.~\ref{tab:summary_stats}. We, therefore, choose to utilize a more conservative upper limit by applying the Minkowsi inequality directly to Eq.~\eqref{eq:combination_sys}:
\begin{equation}
 \sigma_{\rm joint, sys} \leq \sigma_{\rm joint} + \sigma_{\rm sys} + \sqrt{2 \sigma_{\rm sys} \sigma_{\rm joint}} \, .
 \label{eq:upper_bound_new}
\end{equation}

We recommend the right hand side of Eq.~\ref{eq:upper_bound_new} as a conservative prior width for the HSC Y3 cosmological weak lensing analysis. However we strongly recommend performing a sensitivity study for this prior width especially for Bin~4. \referee{We refer to \citet{2023arXiv230400701D}, \citet{2023arXiv230400702L}, \citet{2023arXiv230400703M}, \citet{2023arXiv230400704M} and \citet{2023arXiv230400705S} for further details on the conclusions of this analysis and their implications on prior choices}.

Tab.~\ref{tab:summary_stats} summarizes our results by giving the mean and standard deviation of the posterior mean for the various analysis scenarios presented in this work. The columns list the corresponding results for the Y1 analysis in \citet{2020PASJ...72...16H}, the results obtained for HSC Y3 using the photometry alone with cosmic variance correction (`PhotZ (DNNz)' in Fig.~\ref{fig:joint_inference}), the results we obtain using the DEMPz code and the joint constraints that include the
cross-correlation data vector (`PhotZ \& WX' in Fig.~\ref{fig:joint_inference}). The final column lists the final result that includes the conservative assessment of model error following Eq.~\eqref{eq:upper_bound_new}. 

The error budget we obtain from a combination of cross-correlations and photometry without the additional systematic uncertainty term is almost an order of magnitude smaller than in the HSC Y1 results. The $p_{\rm samp}(z)$ constraints we obtain from the cross-correlation measurements and the $\vec{p}_{\rm indiv}(z)$ are consistent. The model error assessment that we use for our final recommendation on priors is therefore very conservative and is very similar and/or more conservative compared with the error budget in the HSC Y1 analysis. We note that the error budget is dominated by our assessment of model error, i.e., derived by the comparison with the DEMPz method. This assessment of model error is conservative, since the joint constraint between the CAMIRA LRG and the photometry based inference would allow for almost an order of magnitude smaller error in the posterior mean. 

However it is not overly pessimistic and is less than double the residual systematic expected from the difference between the PhotZ (DNNz) and PhotZ\&WX results presented in  Bin~1 of Fig.~\ref{fig:joint_inference_post_mean}. Future work would benefit from adding additional constraints to the high redshift tomographic bin, e.g., by including spatial cross-correlations with DESI and by reconsidering the low redshift systematics in the cross-correlation constraints.

\section{Summary}
\label{sec:discussion}
This work presents posterior sample redshift distributions ($p_{\rm samp}(z)$) in four tomographic bins for the HSC three-year shape catalog. To exploit the synergy between complementary sources of redshift information, we combined $p_{\rm samp}(z)$ constraints from spatial cross-correlations and from individual galaxy photometric redshift distributions ($\vec{p}_{\rm indiv}(z)$) derived from the galaxies photometry. We perform cross-correlation based $p_{\rm samp}(z)$ inference using the CAMIRA Luminous Red Galaxy (LRG) sample, which allowed us to obtain constraints within the limited redshift range of the LRG sample of $z \leq 1.2$. The presented analysis had to account for three main sources of systematic biases and uncertainties: the intrinsic photometric redshift error in the LRGs, the significant variation (both methodologically and in quality) of the provided $\vec{p}_{\rm indiv}(z)$, and the spatial color-redshift dependent selection functions of our specXphot redshift calibration sample.

The goals of the analysis were to provide posteriors for the relevant tomographic $p_{\rm samp}(z)$, demonstrate consistency between the constraints derived using the spatial cross-correlations and $\vec{p}_{\rm indiv}(z)$, and recommend priors on $p_{\rm samp}(z)$ parameters for the cosmological weak lensing (WL) analysis. The latter should also incorporate an assessment of model error and should reflect conservative analysis choices under acceptable degradation of cosmological parameter constraints. We claim that these analysis goals were accomplished in our analysis.

Our analysis was structured as follows (see \S~\ref{sec:photometric_redshift_inference_pipeline}):
\begin{itemize}
    \item Sample definition and selection (\S~\ref{subsec:sample_selection})
    \item Estimation of individual and tomographic $p_{\rm samp}(z)$ using photometry-based inference (Phot, \S~\ref{indiv_galaxy_pz_estimation})
    \item Incorporation of cosmic variance from the spatially limited specXphot training sample into the constraint (\S~\ref{subsec:sample_pz_PhotZ})
    \item Cross-Correlation-based $p_{\rm samp}(z)$ inference (WX, \S~\ref{subsec:joint_inference_camira})
    \item Joint inference combining WX and Phot (\S~\ref{subsec:joint_constraint})
    \item Recommendation of the science-ready photometric redshift priors for WL (\S~\ref{subsec:prior_recommendation_wl})
\end{itemize}
The sample was limited to galaxies with single-peaked $\vec{p}_{\rm indiv}(z)$. The removal of galaxies that show secondary, high redshift ($z > 1.2$) photometric redshift solutions is essential for our analysis, to ensure that we can validate our photometric redshifts with the data products available. Since the CAMIRA LRG sample does not allow a calibration to $z > 1.2$ and the specXphot calibration sample is expected to be incomplete at the faint end of the color-magnitude space, we cannot reliably validate secondary solutions at $z \gtrapprox 1.2$.

This work introduces a framework for sample redshift inference for both empirical methods based on conditional density estimation and methods that are based on SED fitting or likelihood-based forward modelling. Initially we considered three methods for $\vec{p}_{\rm indiv}(z)$ estimation: a likelihood based SED fitting code (Mizuki) and two empirical methods (DNNz, DEMPz).
 
We selected the DNNz method, a conditional density estimation method for photometric redshifts, as our fiducial inference method based on initial comparisons with the cross-correlation data vector. As the specXphot calibration sample used for training the individual galaxy redshift estimators at the faint end of the sample covers only a small solid angle, we construct a logistic Gaussian Process model to parametrize the cosmic variance component in the error model for the inferred tomographic $p_{\rm samp}(z)$.

In the next analysis step, we measured spatial cross-correlations between the CAMIRA LRG and the HSC Y3 photometric shape catalog (HSC phot) for the first three tomographic bins (within $z < 1.2$) and account for the photometric redshift error in the CAMIRA LRG sample in the construction of the cross-correlation likelihood. We demonstrated consistency between the $p_{\rm samp}(z)$ constraints derived from the cross-correlation data vector and photometry-based sample redshift inference.

Utilizing a joint inference framework that accounts for the limited redshift coverage of the cross-correlation measurements, we obtained posterior $p_{\rm samp}(z)$ in four tomographic bins.

Finally we included a conservative error assessment based on a comparison with an alternative photometric redshift algorithm, `DEMPz'. While the final constraint on the mean of the tomographic bins is much narrower than the results obtained in the HSC Y1 analysis \citep{2020PASJ...72...16H}, our conservative assessment of model error yields a prior recommendation for the HSC three-year WL analysis that is similar to (and more conservative than) the Y1 HSC cosmological weak lensing analysis.

\section{Discussion and Future Work}
\label{sec:discussion_futurework}
In the following, we describe a range of known limitations in our analysis that motivate our conservative error assessment and highlight avenues for future work. We concentrate on five areas of this analysis where we identified limitations:
\begin{enumerate}
    \item Error quantification of $\vec{p}_{\rm indiv}(z)$
    \item Treatment of selection functions of the specXphot calibration sample
    \item Treatment of cosmic variance induced by redshift calibration using the specXphot calibration sample
    \item Photometric redshift uncertainties and systematics of CAMIRA LRG galaxies
    \item Simplistic treatment of astrophysical effects in the modeling of the cross-correlation data vector.
\end{enumerate}
In the following paragraphs we will discuss each of these items in order.

(i) There are a number of unmodelled systematics in the construction of $\vec{p}_{\rm indiv}(z)$ using DNNz, DEMPz and Mizuki that are likely explanations for the large differences between their estimates relative to the statistical uncertainty. We show this in Tab.~\ref{tab:summary_stats} where the model error from differences in the DNNz and DEMPz results dominates the error budget. This is qualitatively consistent with the first year HSC analysis\footnote{\citet{2018PASJ...70S...9T} analyze Y1 data. The paper does not present a principled inference strategy to derive $p_{\rm samp}(z)$ for, e.g., Mizuki that requires deconvolving for photometric redshift error (see \S~\ref{subsec:forward_modelling}). However, this does not invalidate a qualitative comparison with our analysis.} of individual galaxy redshift distribution systematics in \citet{2018PASJ...70S...9T}. Fig.~11 and Fig.~14 in that paper illustrate significant differences between the estimates obtained using different methodologies both in terms of the estimated $p_{\rm samp}(z)$ (Fig.~11) and in terms of the PIT metric (Fig.~14), which quantifies how well the $\vec{p}_{\rm indiv}(z)$ are calibrated with respect to a specXphot reference dataset. The significant differences between the methods imply an incomplete assessment of model error\footnote{Model error refers here to error contributions (both systematic and statistical), for example from lack of training data, uncorrected selection functions in the training data, inaccurate modeling of SEDs, priors or photometry.}.

(ii) While the specXphot calibration data were assembled to reduce the impact of unwanted selection functions and we employ the calibration cut (see \S~\ref{subsec:sample_selection}) to remove problematic regions in color space with doubly-peaked $p_{\rm indiv}(z)$, it likely does not provide an unbiased source of redshift calibration for model evaluation and training. Our analysis, therefore, used cross-correlations with the CAMIRA LRG sample, within the aforementioned limited redshift coverage, for redshift calibration and imposed a conservative assessment of model error. The latter is motivated by an acceptable degradation in the cosmological parameter constraints forecasted for the upcoming WL analysis. However future analyses with the full HSC survey dataset and upcoming surveys such as LSST will have to continue to further improve the analysis methodology to reduce this source of systematic uncertainty.

(iii) Our approach to quantify cosmic variance from the spatially small calibration field suffers from three main limitations that we discuss in the following.  We note, however, that the current analysis will likely not be methodologically limited in this area as the dominant source of uncertainty is the model error in the $\vec{p}_{\rm indiv}(z)$. The modelling of the variance of the point field within a patch on the sky depends not only on the point-field expected number density per area and redshift, which can be scaled to match the color-redshift distribution of the target field, but also on the clustering of the galaxies of the underlying process. The latter is modelled based on the COSMOS2015 field, which covers a small area, has different clustering properties than other fields and might be subject to a non-random spatial selection function\footnote{For example, randomly selecting multiple spatially small patches on the sky would show different clustering properties than `favoring interesting' regions with an abundance of clusters and therefore produce a different cosmic-variance model.  }. The small area and nonrandom selection function implies that any statistic derived from this field will not be fully representative of other fields. This means that our cosmic variance estimate derived on COSMOS2015 is not necessarily representative of the true cosmic variance contribution of photometric redshift estimates trained on any COSMOS2015-size patch on the sky. Since the galaxy field is ergodic, this becomes less of a concern for spatially larger fields or if several small but spatially separated fields are used. Furthermore, since the variance does not uniquely identify the stochastic process that describes the $p_{\rm samp}(z)$ uncertainty, every assessment of cosmic variance has model assumptions. We discuss this point in detail in Appendix~\ref{sec:PriorChoice}. We note that we neglect spatial correlations between the COSMOS2015 field and HSC phot, i.e., we do not formulate a full spatial model for redshift inference in this work, which can affect our assessment of cosmic variance. These limitations affect the redshift calibration in other surveys such as DES, which is also based on spatially small calibration fields. We also note that the individual galaxy redshift estimates presented in this work do not allow us to construct a direct relation to the COSMOS2015 training set galaxies, which limits our ability to perform a cosmic variance correction in color space. In future work, we will present a spatial model for redshift inference that will extend the current approach to treat cosmic variance in $p_{\rm samp}(z)$ estimation (Rau et al. in prep.).

(iv) Our modelling of the WX data vector depends on accurately parametrizing the photometric redshift systematics of the CAMIRA LRG sample. As discussed, especially at low redshift these systematics can be quite significant. Our current modelling is based on a specXphot calibration sample, as we did not obtain access to the relevant CAMIRA LRG likelihoods. As a result, our correction could be subject to residual systematics from spectroscopic selection functions. This needs to be reconsidered in the future, along with a better assessment of galaxy-dark matter bias for the calibration sample. This includes parametrizing a redshift and scale dependence in the galaxy-dark matter bias  within each tomographic bin for the photometric sample and the calibration sample. In order to constrain this more complex assessment of galaxy-dark matter bias, it will be important to extend the data vector towards auto-correlations of the photometric and reference samples. 

(v) Regarding the modelling of the cross-correlation data vector, we limited our analysis to a constant galaxy-dark matter bias within each tomographic source bin and did not include an assessment of magnification bias.  \citet{2022MNRAS.510.1223G} studied the effect of magnification bias on cross-correlation based $p_{\rm samp}(z)$ inference in the context of the Dark-Energy-Survey Year 3 analysis. While performed in the context of different data and analysis, we can expect the effect of magnification bias to be subdominant compared with the modelling of a redshift-dependent galaxy-dark matter bias and subdominant compared with our conservative total error budget. While based on a qualitative extrapolation of their quantitative assessments \citep[see Tab.~2][]{2022MNRAS.510.1223G}, the good agreement between WX and Y3 PhotZ reported in Fig.~\ref{fig:joint_inference} provide some basis for that claim. Future measurements with larger signal-to-noise will need to reconsider this assumption. 

In conclusion, we have presented a $p_{\rm samp}(z)$ inference methodology for the HSC Y3 shape catalog that represents a significant update over the methodology in previous HSC weak lensing analyses. We have forecasted the effect of our updated methodology on the previous HSC S16A analysis in \S~\ref{sec:impact_s16a_cosmology} and demonstrated that our updated methodology can account for shifts in the $\Omega_m$-$S_{8}$ plane of $0.5 \sigma$ after rescaling the covariance matrix from previous HSC weak lensing measurements to account for the increased area in the HSC Y3 catalog. This highlights the importance of sample redshift calibration as we prepare not only for the HSC analysis but also look ahead towards upcoming surveys like LSST.

\section*{Acknowledgements}
\referee{We thank the anonymous referee for helpful comments that improved both content and presentation of the paper. The HSC Collaboration acknowledges fundamental work on photometric redshifts by the Complete Calibration of the Color Redshift Relation (C3R2) team.}
The Hyper Suprime-Cam (HSC) collaboration includes the astronomical communities of Japan and Taiwan, and Princeton University. The HSC instrumentation and software were developed by the National Astronomical Observatory of Japan (NAOJ), the Kavli Institute for the Physics and Mathematics of the Universe (Kavli IPMU), the University of Tokyo, the High Energy Accelerator Research Organization (KEK), the Academia Sinica Institute for Astronomy and Astrophysics in Taiwan (ASIAA), and Princeton University. Funding was contributed by the FIRST program from the Japanese Cabinet Office, the Ministry of Education, Culture, Sports, Science and Technology (MEXT), the Japan Society for the Promotion of Science (JSPS), Japan Science and Technology Agency (JST), the Toray Science Foundation, NAOJ, Kavli IPMU, KEK, ASIAA, and Princeton University.

This paper makes use of software developed for Vera C. Rubin Observatory. We thank the Rubin Observatory for making their code available as free software at http://pipelines.lsst.io/.

This paper is based on data collected at the Subaru Telescope and retrieved from the HSC data archive system, which is operated by the Subaru Telescope and Astronomy Data Center (ADC) at NAOJ. Data analysis was in part carried out with the cooperation of Center for Computational Astrophysics (CfCA), NAOJ. We are honored and grateful for the opportunity of observing the Universe from Maunakea, which has the cultural, historical and natural significance in Hawaii.

The Pan-STARRS1 Surveys (PS1) and the PS1 public science archive have been made possible through contributions by the Institute for Astronomy, the University of Hawaii, the Pan-STARRS Project Office, the Max Planck Society and its participating institutes, the Max Planck Institute for Astronomy, Heidelberg, and the Max Planck Institute for Extraterrestrial Physics, Garching, The Johns Hopkins University, Durham University, the University of Edinburgh, the Queen’s University Belfast, the Harvard-Smithsonian Center for Astrophysics, the Las Cumbres Observatory Global Telescope Network Incorporated, the National Central University of Taiwan, the Space Telescope Science Institute, the National Aeronautics and Space Administration under grant No. NNX08AR22G issued through the Planetary Science Division of the NASA Science Mission Directorate, the National Science Foundation grant No. AST-1238877, the University of Maryland, Eotvos Lorand University (ELTE), the Los Alamos National Laboratory, and the Gordon and Betty Moore Foundation.

Work at Argonne National Laboratory was supported by the U.S. Department of Energy, Office of High Energy Physics. Argonne, a U.S. Department of Energy Office of Science Laboratory, is operated by UChicago Argonne LLC under contract no. DE-AC02-06CH11357. MMR acknowledges the Laboratory Directed Research and Development (LDRD) funding from Argonne National Laboratory, provided by the Director, Office of Science, of the U.S. Department of Energy under Contract No. DE-AC02-06CH11357.
MMR’s work at Argonne National Laboratory was also supported
under the U.S. Department of Energy contract DE-AC02-06CH11357.
RD acknowledges support from the NSF Graduate Research Fellowship Program under Grant No.\ DGE-2039656. Any opinions, findings, and conclusions or recommendations expressed in this material are those of the authors and do not necessarily reflect the views of the National Science Foundation. AJN is supported by Grant-in-Aid for Transformative Research Areas 21H05454, and JSPS KAKENHI Grant Numbers JP20H0193 and JP21K03625.
RM is supported by DOE grant DE-SC0010118 and a grant from the Simons Foundation (Simons Investigator in Astrophysics, Award ID 620789).
MT is supported by World Premier International Research Center Initiative (WPI Initiative),  JSPS KAKENHI Grant Numbers JP20H05850, JP20H05855,
and JP19H00677 and by Basic Research Grant (Super AI) of Institute for AI and Beyond of the University of Tokyo. 

%%%%%%%%%%%%%%%%%%%%%%%%%%%%%%%%%%%%%%%%%%%%%%%%%%
\section*{Data Availability}
This work is part of the HSC Year 3 cosmological analysis. The data and analysis products, as well as the software, will be made publicly available via the HSC-SSP website \url{https://hsc.mtk.nao.ac.jp/ssp/data-release/}.

% The inclusion of a Data Availability Statement is a requirement for articles published in MNRAS. Data Availability Statements provide a standardised format for readers to understand the availability of data underlying the research results described in the article. The statement may refer to original data generated in the course of the study or to third-party data analysed in the article. The statement should describe and provide means of access, where possible, by linking to the data or providing the required accession numbers for the relevant databases or DOIs.



\bibliographystyle{mnras}
\bibliography{example} % if your bibtex file is called example.bib

%\begin{thebibliography}{99}
%\end{thebibliography}

%%%%%%%%%%%%%%%%%%%%%%%%%%%%%%%%%%%%%%%%%%%%%%%%%%

%%%%%%%%%%%%%%%%% APPENDICES %%%%%%%%%%%%%%%%%%%%%

\appendix

\section{Conditional density estimation methodology}
\label{sec:methodology_direct_mapping}
\paragraph*{Overview}
In the following appendix, we describe our methodology to perform sample redshift inference in the context of conditional density estimation in continuation of \S~\ref{subsec:conddens}. The discussion in this section applies to the DNNz and DEMPz methods. The basic idea of conditional density redshift estimation is to construct an estimator of the true conditional density $p(z | \mathbf{f})$ of the redshift $z$ given the fluxes (or photometry) $\mathbf{f}$. We construct this mapping between the `true' redshift $z$ and measured flux, which requires a specXphot `training' dataset. This can be constructed using spatially overlapping spectroscopic and photometric survey data, which provides both photometry and accurate spectroscopic redshifts. Upon constructing a conditional density estimator $\hat{p}(z | \mathbf{f})$, for a particular photometric survey, we can construct an estimator of the $p_{\rm samp}(z)$ as
\begin{equation}
    \hat{p}_{\rm samp}(z) = \int \hat{p}(z | \mathbf{f}) \hat{p}(\mathbf{f}) {\rm d}\mathbf{f} \, ,
    \label{eq:marginalized_conddens}
\end{equation}
where $\hat{p}(z | \mathbf{f})$ and $\hat{p}(\mathbf{f})$ denote estimators of the conditional density of redshift $z$ given flux $\mathbf{f}$ and of the marginal density of color-magnitude space $\hat{p}(\mathbf{f})$. We note here the difference between constructing an estimator of the conditional density $\hat{p}(z | \mathbf{f})$ and a `forward modeling' approach that would require the formulation of a likelihood (or the nonparametric estimator thereof) $p(\mathbf{f} | z)$. The former is a density estimation problem and requires the availability of a calibration dataset to provide information on the redshift distribution of galaxies as a function of measured photometry. The latter induces an inverse problem that depends on knowledge of the data-generating process from a true redshift to measured photometry. One would include additional redshift information here in the formulation of the prior. We stress that these model formulations are very different and refer to \S~\ref{sec:derivation_inference_scheme} for a detailed description of the redshift inference methodology in the context of likelihood-based forward modeling. In the following, we will comment on the assumptions behind the conditional density estimation methodology.
\paragraph*{Assumptions}
The basic assumption of empirical methods is that the data in the calibration and target datasets follow the same conditional densities $p(z | \mathbf{f})$ of the redshift $z$ given the fluxes (or photometry) $\mathbf{f}$. We also note that there exist no unbiased nonparametric density estimators \citep{10.1214/aoms/1177728190}. Therefore, a prime challenge for these methods is the selection of bandwidth, or smoothing scale.

Inaccurate selection of smoothing can lead to biases in redshift inference that are relevant for cosmological inference, as shown in prior work \citep{2017MNRAS.466.2927R}. The authors also demonstrated, that biases from inaccurate selection of smoothing can be mitigated in cosmological inference using parametric bootstrap techniques. In the context of this work, we can assume that misspecification errors due to non-representative training data and epistemic uncertainty from a lack of training data will be more severe than biases due to inaccurate bandwidth selection.
\paragraph*{Methodology}
In the following, we construct an empirical likelihood of the density estimator Eq.~\ref{eq:marginalized_conddens} that allows us to marginalize over systematics in a principled way.
Under the assumptions described in the previous paragraph, we can parameterize $p(z | \mathbf{f})$ as a histogram
    \begin{equation}
        p(z | \mathbf{f}) = \sum_{i = 1}^{N_{\rm bins}} w_i(\mathbf{f}) \mathds{1}_i(z) \, ,
    \end{equation}
where $w_i$ denotes the histogram bin height and $\mathds{1}_i(z)$ is unity if the redshift is within bin $i$, and zero otherwise. $N_{\rm bins}$ denotes the number of histogram bins.

This yields an estimator for $p(z)$ as
\begin{equation}
    p(z) = \sum_{i = 1}^{N_{\rm bins}} \left(\int w_i(\mathbf{f}) p(\mathbf{f}) {\rm d}\mathbf{f}\right) \mathds{1}_i(z) = \sum_{i = 1}^{N_{\rm bins}} E_{\mathbf{f}}\left[ w_i(\mathbf{f}) \right] \mathds{1}_i(z)  \, ,
    \label{eq:stacked_pz_conddens}
\end{equation}
where $E_{\mathbf{f}}\left[ w_i(\mathbf{f}) \right]$ denotes the expectation value of the weights $w_i(\mathbf{f})$ wrt. to the marginal distribution of photometry. The weights $w_i(\mathbf{f})$ can depend on parameters $\boldsymbol{\eta}$ that describe additional sources of error, induced by unmodelled selection functions in the training data or by intrinsic model bias in the conditional density estimates.

Based on this relation, we can employ the empirical likelihood formalism \citep[e.g.,][]{10.1214/aos/1176347494, owen2001empirical, pawitan2001all} and construct an estimating equation
\begin{equation}
    \boldsymbol{\psi}([\boldsymbol{\phi_\text{nz}}, \boldsymbol{\eta}], \mathbf{f}) = \boldsymbol{w}(\mathbf{f}, \boldsymbol{\eta}) - \boldsymbol{\phi_\text{nz}}
\end{equation}
where $\boldsymbol{\phi_\text{nz}}$ denotes the modelled histogram heights (see Eq.~\eqref{eq:parametrization_pz}) of the $p_{\rm samp}(z)$. We note here that $\boldsymbol{w}(\mathbf{f}, \boldsymbol{\eta})$ is a function of the measured photometry and parameters that describe other systematics, whereas $\boldsymbol{\phi_\text{nz}}$ is the parameter vector to be estimated.

Under the assumption that the parameter set $\boldsymbol{\eta}$ accurately describes the systematics mentioned above, we seek values for $\boldsymbol{\eta}$ and $\boldsymbol{\phi_\text{nz}}$ such that
\begin{equation}
    E_{\mathbf{f}}\left[\boldsymbol{\psi}([\boldsymbol{\phi_\text{nz}}, \boldsymbol{\eta}], \mathbf{f}) \right] = 0 \, .
\end{equation}
We can treat the application of lensing weights $\omega_{\rm lens}(\mathbf{f})$ as a selection function and follow the recipe described in \citet{owen2001empirical} of modifying the expected estimating equation by transforming the probability measure as
\begin{equation}
0 = \int \boldsymbol{\psi}([\boldsymbol{\phi_\text{nz}}, \boldsymbol{\eta}], \mathbf{f}) dF(\mathbf{f}) = \int \boldsymbol{\psi}_{\rm WL}([\boldsymbol{\phi_\text{nz}}, \boldsymbol{\eta}], \mathbf{f})\omega_{\rm lens}(\mathbf{f}) dG(\mathbf{f}) \, ,
\end{equation}
where $\omega_{\rm lens}(\mathbf{f})$ denotes the lensing weights as a function of photometry (and other auxillary parameters omitted here). In the following we will \referee{omit} the dependence of the lensing weights on $\mathbf{f}$ for convenience. It is understood that the introduction of lensing weights implies a dependence on a variety of parameters that describe the measurement of galaxy shapes.

($dF(\mathbf{f})$/$dG(\mathbf{f})$) denotes the (unweighted/weighted) probability measures where $dF(\mathbf{f}) = \omega_{\rm lens} dG(\mathbf{f})$.

We introduce ($\boldsymbol{w_{\rm WL}}(\mathbf{f}, \boldsymbol{\eta})$/$\boldsymbol{\phi_\text{nz}}_{\rm WL}$) that denote the weighted (measured/modeled) weak-lensing histogram height parameters that include lensing weights as
\begin{equation}
    \begin{split}
        \boldsymbol{w_{\rm WL}}(\mathbf{f}, \boldsymbol{\eta}) &= \boldsymbol{w}(\mathbf{f}, \boldsymbol{\eta}) \, \omega_{\rm lens} \\
       \boldsymbol{\phi_\text{nz}}_{\rm WL} &=  \boldsymbol{\phi_\text{nz}} \,  \omega_{\rm lens} \, .
    \end{split}
\end{equation}
The new estimating equation $\boldsymbol{\psi}_{\rm WL}([\boldsymbol{\phi_\text{nz}}_{\rm WL}, \boldsymbol{\eta}], \mathbf{f})$ is now adjusted for the lensing weights and can be used in conjunction with the empirical likelihood framework to define a likelihood on the mean $E_{\mathbf{f}}\left[ w_{\rm i, WL}(\mathbf{f}, \boldsymbol{\eta}) \right]$ in Eq.~\ref{eq:stacked_pz_conddens}. We reiterate that $E_{\mathbf{f}}\left[ w_{\rm i, WL}(\mathbf{f}, \boldsymbol{\eta}) \right]$ denotes here the expectation over the $w_{\rm i, WL}(\mathbf{f}, \boldsymbol{\eta})$ corresponding to bin $i$ over all galaxies in the sample.

The empirical likelihood framework is a nonparametric approach to estimation, that imposes an empirical discrete distribution over the weights $\boldsymbol{w_{\rm WL}}(\mathbf{f}, \boldsymbol{\eta})$ and then utilizes Lagrange multipliers to constrain this distribution such that the discrete probabilities sum to unity, are positive, and the estimating function relation
\begin{equation}
    E_{\mathbf{f}}\left[\boldsymbol{\psi}_{\rm WL}([\boldsymbol{\phi_\text{nz}}_{\rm WL}, \boldsymbol{\eta}], \mathbf{f})\right] = 0
\end{equation}
is full-filled. One can show in analogy to \citet{owen2001empirical} that a profile log-likelihood on the mean Eq.~\ref{eq:stacked_pz_conddens} is obtained by finding the roots to
\begin{equation}
    g(\boldsymbol{\lambda}) = \sum_{i = 1}^{N_{\rm gal}} \left(\frac{\boldsymbol{w}_{\rm WL}(\mathbf{f}_i, \boldsymbol{\eta}) - \boldsymbol{\phi_\text{nz}}_{\rm WL}}{N_{\rm gal} - \boldsymbol{\lambda}^{T} \left(\boldsymbol{w}_{\rm WL}(\mathbf{f_i}, \boldsymbol{\eta}) - \boldsymbol{\phi_\text{nz}}_{\rm WL}\right)} \right) \, ,
    \label{eq:emp_like}
\end{equation}
and subsequently evaluating the profile log-likelihood as
\begin{equation}
    \ell([\boldsymbol{\eta}, \boldsymbol{\phi_\text{nz}}_{\rm WL}]) = - \sum_{i = 1}^{N_{\rm gal}} \log{\left(N_{\rm gal} - \boldsymbol{\lambda}^{T} (\boldsymbol{w_{\rm WL}}(\mathbf{f}_{i}, \boldsymbol{\eta}) - \boldsymbol{\phi_\text{nz}}_{\rm WL})\right)} \, .
    \label{eq:profile_like_emp}
\end{equation}

Eq.~\eqref{eq:emp_like} is monotonic in $\boldsymbol{\lambda}$, which is a Lagrange multiplier of dimension $N_{\rm bins}$. Here, $N_{\rm gal}$ denotes the number of galaxies in the sample. We reach a root for $\boldsymbol{\lambda} = 0$, where $\boldsymbol{\phi_\text{nz}}_{\rm WL} = \frac{1}{N_{\rm gal}} \sum_{i = 1}^{N_{\rm gal}} \boldsymbol{w}_{\rm WL}(\mathbf{f}_i, \boldsymbol{\eta})$. This corresponds to the empirical mean of the weights $\boldsymbol{w}_{\rm WL}(\mathbf{f}, \boldsymbol{\eta})$, often referred to as the `stacked distribution.' This terminology is conventional but misleading because it is often applied inappropriately to summing up likelihood functions of forward models, which is an undefined operation. We refer to \S~\ref{sec:derivation_inference_scheme} for a discussion on estimating the $p_{\rm samp}(z)$ in this context.

The central limit theorem holds for the empirical likelihood framework and the coverage error converges as $1 \big/ N$, where $N$ denotes the sample size \citep{owen2001empirical}. Thus, for the large sample sizes considered in this work, we can safely neglect the statistical error in the maximum empirical likelihood estimate, given that other error contributions, such as model misspecification error and cosmic variance, are considerably larger.

\section{ Forward Modelling Methodology}
\label{sec:derivation_inference_scheme}
\paragraph*{Overview}
In this appendix, we describe the forward modeling formulation of sample redshift inference in more detail and derive a variational inference scheme to perform efficient $p_{\rm samp}(z)$ inference in this framework. In \S~\ref{subsec:forward_modelling} we discussed a simplified model, focussing on the redshift $z$ as the quantity of interest, as \citep[e.g.][]{2016MNRAS.460.4258L, 2020arXiv200712178M, 2022MNRAS.509.4886R}
\begin{equation}
    p(\mathbf{\hat{F}} | \boldsymbol{\phi_\text{nz}}, \boldsymbol{\Omega}) = \prod_{i = 1}^{N_{\rm gal}} \int \mathrm{d}z_i \, \omega_i \, p(\mathbf{f}_i | z_i, \boldsymbol{\Omega}) p(z_i | \boldsymbol{\phi_\text{nz}}, \boldsymbol{\Omega}) \, .
    \label{eq:forward_modelling}
\end{equation}
We reiterate, $\mathbf{\hat{F}}$ denotes the set of fluxes of all $N_{\rm gal}$ galaxies in the sample, $\mathbf{f}_i\,(z_i)$ denotes the flux (redshift) of the individual galaxy with index $i$, and $\boldsymbol{\Omega}$ denotes a set of auxiliary parameters that describe other galaxy properties such as galaxy type or stellar mass. The weights $\omega_i$ denote the lensing weights for each galaxy in the sample. 
%To simplify the notation, we will implicitly assume conditioning on $\boldsymbol{\Omega}$ but omit it from the notation in the following discussion.
\paragraph*{Assumptions}
The simplified Eq.~\eqref{eq:forward_modelling} assumes that the flux and redshift of each galaxy are drawn independently of any other. In a more general setting, we could formulate a joint likelihood. The forward modeling approach does not assume the availability of calibration data and is, therefore, more general than the conditional density estimation methodology. In contrast to conditional density estimation, Eq.~\ref{eq:forward_modelling} implies a hierarchical inference of the $p_{\rm samp}(z)$. The same applies to other population distributions for quantities of interest. For noisy measurements of photometry, this inverse problem can be poorly conditioned. Practical applications must impose explicit or implicit assumptions to control the posterior variance, either by setting priors on quantities of interest or restricting the complexity of relevant models. Model misspecification error is a significant complication in this context. Given the complex modeling of SEDs, selection functions, and photometric error, any practical application must verify their modeling assumptions on calibration data.
\paragraph*{Methodology}
We discretize the $\vec{p}_{\rm indiv}(z)$ on the same grid that defines the $p_{\rm samp}(z)$ histogram defined in Eq.~\eqref{eq:parametrization_pz}. We define a matrix defined as the set:
\begin{equation}
    \boldsymbol{\rm pz} := \{ pz_{\rm i j}(\boldsymbol{\Omega}) \, | \, 0 < i \leq N_{\rm gal}, 0 < j \leq N_{\rm bins}  \} \, ,
\end{equation}
where the entries are given as the integrals of the likelihood of galaxy $i$ over the $j$ redshift histogram bin weighted by the lensing weights $\omega_i$
\begin{equation}
    pz_{\rm i j}(\boldsymbol{\Omega}) := \omega_i \int p(\mathbf{f}_i | z_i, \boldsymbol{\Omega}) \mathds{1}_j(z) \mathrm{d}z_i \, .
\end{equation}

Using the definition Eq.~\eqref{eq:parametrization_pz} we can write the log-likelihood as
\begin{equation}
     \log{\left(p(\mathbf{\hat{F}} | \boldsymbol{\phi_\text{nz}}, \boldsymbol{\Omega})\right)} = \sum_{i = 1}^{N_{\rm gal}} \log{\left( \sum_j^{N_{\rm bins}} \phi_{\rm nz, j} pz_{\rm i j}(\boldsymbol{\Omega}) \right)}  \, .
     \label{eq:def_loglikelihood}
\end{equation}
The logarithm in Eq.~\eqref{eq:def_loglikelihood} and the fact that $\boldsymbol{\phi_\text{nz}}$ is normalized (it can be transformed to lie on the simplex) makes the evaluation and optimization of Eq.~\eqref{eq:def_loglikelihood} nontrivial\footnote{A possible way to perform the optimization in a brute-force approach is by projected gradient descent. However, we derive a simpler scheme in the following. }.

We can circumvent both issues by introducing the binary variables $\rho_{ij}$ that associate bin $j$ with galaxy $i$. The complete data likelihood then reads
\begin{equation}
    p(\hat{\mathbf{F}}, \boldsymbol{\rho} | \boldsymbol{\phi_\text{nz}}, \boldsymbol{\Omega}) \propto \prod_{i = 1}^{N_{\rm gal}} \prod_{j = 1}^{N_{\rm bins}} \left(\phi_{\rm nz, j} \, pz_{\rm i j}(\boldsymbol{\Omega})\right)^{\rho_{\rm i j}} \,
\end{equation}
that we identify as a multinominal likelihood. Imposing a Dirichlet prior over the parameters $\boldsymbol{\phi_\text{nz}}$ then yields the joint distribution $p(\mathbf{\hat{F}}, \boldsymbol{\rho}, \boldsymbol{\phi_\text{nz}} | \boldsymbol{\Omega})$. 
%In the following, we will omit the variable $\boldsymbol{\Omega}$ to simplify the notation.

Variational inference maximizes the Evidence Lower Bound (ELBO), which is equivalent to minimizing the Kullback-Leibler divergence between the true, unknown, posterior, and an `ansatz,' the variational distribution \referee{
\begin{equation}
{\rm ELBO}  = E_{q(\boldsymbol{\rho}, \boldsymbol{\phi_\text{nz}})} \left[ \log{\left( p(\mathbf{\hat{F}}, \boldsymbol{\rho}, \boldsymbol{\phi_\text{nz}} | \boldsymbol{\Omega}) \right) - \log{q(\boldsymbol{\rho}, \boldsymbol{\phi_\text{nz}})} }\right] \, ,
\end{equation}}
where $q(\boldsymbol{\rho}, \boldsymbol{\phi_\text{nz}})$ denotes the variational distribution to be optimized. \referee{Here, this involves imposing an analytic form for the variational distribution and then maximizing the {\rm ELBO} with respect to its parameters. } 

We make a mean-field ansatz for the variational distribution
\begin{equation}
    q(\boldsymbol{\rho}, \boldsymbol{\phi_\text{nz}}) \approx q(\boldsymbol{\rho}) q(\boldsymbol{\phi_\text{nz}}) \, ,
\end{equation}
that assumes independence between $\boldsymbol{\rho}$ and $\boldsymbol{\phi_\text{nz}}$.

Under the mean-field approximation, variational inference reduces to a simple scheme of updating each component iteratively by mean-field coordinate ascent. Setting the Lagrange function constructed using the variational derivative of the ELBO to zero, we can derive the following coordinate ascent iteration steps
\begin{equation}
\begin{split}
    q(\boldsymbol{\rho}) &\propto \exp{\left( E_{q(\boldsymbol{\phi_\text{nz}})} \left[\log{p(\boldsymbol{\rho} | \boldsymbol{\phi_\text{nz}}, \mathbf{pz})}\right]\right)} \\
    &\propto \prod_{i = 1}^{N_{\rm gal}} \prod_{j = 1}^{N_{\rm bins}} \left( \exp{\left(\psi(\alpha_j) - \psi\left(\sum_{a = 1}^{N_{\rm bins}} \alpha_a\right) + \log{(pz_{ij})}\right)}\right)^{\rho_{i j}} \, ,
    \end{split}
    \label{eq:definition_var_distri_rho}
\end{equation}
and
\begin{equation}
\begin{split}
    q(\boldsymbol{\phi_\text{nz}}) &\propto \exp{\left( E_{q(\boldsymbol{\rho})} \left[\log{p(\boldsymbol{\phi_\text{nz}} | \boldsymbol{\rho}, \mathbf{pz})}\right]\right)} \\
    & = {\rm Dir}(\boldsymbol{\alpha_0} + \sum_{i = 1}^{N_{\rm gal}} \boldsymbol{\gamma_{i}})\,
    \end{split}
    \label{eq:definition_var_distri_nz}
\end{equation}
where we have omitted the conditioning of the variational distributions on the parameter $\boldsymbol{\alpha}$ for notational convenience. We note that $\boldsymbol{\alpha}$ is iteratively updated in the argument of the Dirichlet defined in Eq.~\ref{eq:definition_var_distri_nz}. The sum in Eq.~\ref{eq:definition_var_distri_nz} goes over the $N_{\rm gal}$-dimension of the matrix, whose elements are defined as
\begin{equation}
    \gamma_{ij} = \frac{\exp{\left(\psi(\alpha_j) - \psi(\sum_{a = 1}^{N_{\rm bins}}{\alpha_a}) + \log{(pz_{\rm ij})}\right)}}{\sum_{j = 1}^{N_{\rm bins}} \exp{\left(\psi(\alpha_j) - \psi(\sum_{a = 1}^{N_{\rm bins}}{\alpha_a}) + \log{(pz_{\rm ij})}\right)}} \, .
\end{equation}
Here, $\psi$ denotes the digamma function, the Dirichlet distribution is abbreviated as `Dir'. The variational distributions defined in Eq.~\ref{eq:definition_var_distri_rho} and Eq.~\ref{eq:definition_var_distri_nz} are iteratively updated until convergence.

While this iterative scheme can be expected to computationally outperform MCMC approaches, a mean-field ansatz often leads to the estimation of too narrow credibility intervals.

In our numerical experience, the under-coverage\footnote{Under-coverage refers here to underestimating the width of the credibility intervals.}  under reasonable regularization (e.g. by selecting broader histogram bins) is approximately 20\%, which is subdominant compared with other sources of error induced by spatial-, color- and redshift-dependent selection functions or model misspecification. We, therefore, used the variational inference scheme in this work during the initial stages of the project, where we evaluated the accuracy of the Mizuki individual galaxy photometric redshifts. However, we note that the validity of the variational inference approximation will depend on the resolution (e.g. given by the histogram bins size) and can be expected to deteriorate for poorly conditioned scenarios with high variance. In these cases, we can expect credibility intervals to exhibit under-coverage. In contrast, maximum \referee{{\em a posteriori}} predictions can be expected to be still entirely accurate.

\section{Characterizing Cosmic Variance using Logistic Gaussian Processes}
\label{sec:characterizing_cv_loggp}
In this appendix we discuss how logistic Gaussian processes provide a flexible model to include cosmic variance induced sample noise into $p_{\rm samp}(z)$ inference.
\begin{figure*}
    \centering
    \includegraphics[scale=0.6]{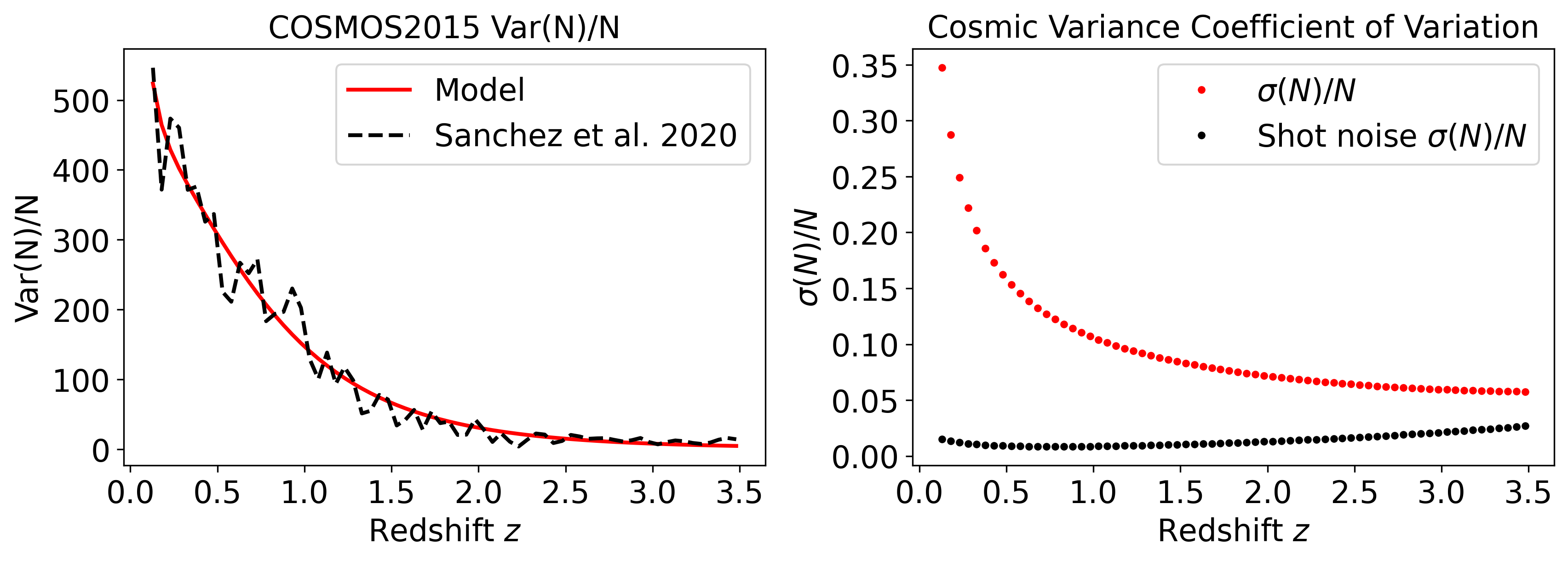}
    \caption{ \textit{Left:} $Var(N)\big/N$ model for the COSMOS2015 data as a function of redshift used in this work (red) compared with the (black, dashed) predictions in \citet{2020arXiv200409542S}. \textit{Right:} Coefficient of variation  as a function of redshift predicted by our model (red) compared with the contribution from shot noise alone (black).  }
    \label{fig:comic_variance}
\end{figure*}

We first consider the redshift-dependent log-normal doubly stochastic point process specified as
\begin{equation}
    \begin{split}
        \rho_i &\sim {\rm LogNorm}(\boldsymbol{\mu}, \boldsymbol{\Sigma}) \\
        N_i &\sim {\rm Poisson}(\rho_i) \, ,
    \end{split}
    \label{eq:cox_process}
\end{equation}
where $\boldsymbol{\mu}$ and $\boldsymbol{\Sigma}$ are the mean and covariance parameters of the lognormal distribution, the $\rho_i$ are the mean parameters of the Poisson distribution that describes the galaxy number counts in redshift dimension and $N_i$ denotes the number of galaxies in each redshift bin. The $p_{\rm samp}(z)$, that enters the modelling of two-point statistics, is normalized to integrate to unity. We therefore need to sample over normalized histogram counts of a multinomial instead of parameters of a Poisson distribution. 

The lognormal `Cox process' defined in Eq.~\eqref{eq:cox_process} can be equivalently defined as
\begin{equation}
        \begin{split}
        \rho_i &\sim {\rm LogNorm}(\boldsymbol{\mu}, \boldsymbol{\Sigma}) \\
        \mathbf{N} &\sim {\rm MultNominal}(\boldsymbol{\phi_\text{nz}}, \overline{N})  \\
        \overline{N} &\sim {\rm Poisson}(\overline{\rho}) \\
       \phi_{{\rm nz}, i} &= \frac{\rho_i}{\sum_{i = 0}^{N_{\rm bins}} \rho_i}
        \\
        \overline{\rho} &= \sum_{i = 0}^{ N_{\rm bins}} \rho_i \, ,
    \end{split}
    \label{eq:logistic_gp}
\end{equation}
where we have decomposed the Poisson distribution into the product of a Multinominal distribution that depends on the normalized $\boldsymbol{\rho}$-parameters and a Poisson distribution that depends on their sum $\overline{\rho}$. The random variable $\overline{N}$ denotes the total number of galaxies across all bins.

Here, $N_{\rm bins}$ is the number of redshift bins introduced in Eq.~\eqref{eq:parametrization_pz}. Since the modelling of the angular correlation function depends on the normalized $p_{\rm samp}(z)$, we will concentrate on the distribution of $\boldsymbol{\phi_\text{nz}}$, where Eq.~\eqref{eq:logistic_gp} defines the logistic Gaussian process specification of our model. We make the simplifying assumption of ignoring the cross-correlations between neighboring redshift bins, which has been shown to be a reasonable approximation in \citet{2020arXiv200409542S}.
To include an error contribution to the lognormal model covariance that matches the variation in the COSMOS field, we are interested to predict the coefficient of variation, i.e. the ratio between the standard deviation and mean, for the HSC phot data in the COSMOS field as a function of redshift due to cosmic variance and use it to derive a cosmic variance error budget on the $p_{\rm samp}(z)$ model for each tomographic bin. To this end, we first formulate a model for the variance of galaxy counts as a function of redshift that can be fitted to the results of \citet{2020arXiv200409542S}.

Consider two sets of galaxies within a spatial area and redshift bin, that we denote as $B_{1}$ and $B_{2}$. We can express the covariance of the number of galaxies within the sets $N(B_{1/2})$ as \citep[e.g.][]{stoyan1994fractals}
\begin{equation}
\begin{split}
    &{\rm cov}(N(B_{1}), N(B_{2})) = E\left[N(B_{1})N(B_{2})\right] - E\left[N(B_{1})\right] E\left[N(B_{2})\right] \\
    &= E\left[\sum_{\mathbf{x_1} \in N(B_{1})} \sum_{\mathbf{x_2} \in N(B_{2})} \mathds{1}_{B_{1}}(\mathbf{x_1}) \mathds{1}_{B_{2}}(\mathbf{x_2})\right] - \rho_{B_1} V(B_1) \rho_{B_2} V(B_2) \, ,
\end{split}
\end{equation}
where $V(B_{1/2})$ and $\rho_{B_{1/2}}$ denotes the volume and expected number density of $B_{1}$ and $B_{2}$. The \referee{volume} is defined with respect to spatial area and redshift bin and the expected number density $\rho$ denotes the expected number of galaxies observed in $B$ per unit volume. $\mathds{1}_B(\mathbf{x})$ denotes the indicator function which is unity if a galaxy can be found at position $\mathbf{x}$ and zero otherwise. The first term corresponds to the second-moment measure, i.e. the expected number of galaxy pairs including `pairs' of the same galaxy. This can be expressed as a function of the two point correlation function, the number densities and the effect of the survey mask. The variance contribution we obtain within a set $B$ under the assumption of homogeneity and isotropy can be defined as
\begin{equation}
    Var[N(B)] = \rho_B V(B) +  \rho_{B}^2 \iint\limits_{B} \xi(||\mathbf{x_1} - \mathbf{x_2}||) d\mathbf{x_1} d\mathbf{x_2} \, .
    \label{eq:variance}
\end{equation}
The first term in Eq.\ref{eq:variance} is the `shot noise' contribution. The second term in Eq.~\eqref{eq:variance} depends on the `clustering' of the galaxy field, parametrized by the pair-correlation function $\xi(x_1, x_2)$ and the survey geometry that enters the double integral over $B$.

We develop a simple model for the COSMOS2015 data based on Eq.~\eqref{eq:variance} by parametrizing $\rho_B$ proportional to a lognormal distribution and the integral of the correlation function proportional to a power law. Our model has 5 parameters; an amplitude and scale parameter for the power law model and two parameters that describe the line-of-sight number density of the COSMOS2015 number counts with a normalization amplitude. We then fit this model to the redshift dependent $Var[N]/N$ values reported in \citet{2020arXiv200409542S}, shown in the left panel of Fig.~\ref{fig:comic_variance}. The black dashed line shows the values reported in \citet{2020arXiv200409542S}, the red line shows the best fit solution to our model. We see that at low redshift the linear dependence on the log-normal shaped line-of-sight number density of the COSMOS2015 number counts flattens the power law shape. In the right panel of Fig.~\ref{fig:comic_variance} we plot the coefficient of variation (red) and the coefficient of variation\footnote{The coefficient of variation is the ratio of the standard deviation to the mean.} from only the shot noise contribution, i.e. the first term of Eq.~\eqref{eq:variance}. In agreement with  \citet{2020arXiv200409542S} we see, that the shot noise contribution is subdominant for the COSMOS2015 dataset. This difference will be even larger for our data due to the larger amount of galaxies in HSC phot.

The cosmic variance contribution to the coefficient of variation is strictly bounded from above by the total coefficient of variation by
\begin{equation}
    \sqrt{\iint\limits_{B} \xi(||\mathbf{x_1} - \mathbf{x_2}||) d\mathbf{x_1} d\mathbf{x_2}} \leq \sigma[N(B)]\big/E[N(B)] \, .
    \label{eq:cosmic_variance_bound}
\end{equation}
We choose to use the `full' coefficient of variation from COSMOS2015 (`CV'), in our model, even though the shot noise contribution would already be included in the empirical likelihood framework (or the deconvolution approach in the Mizuki case), which will lead to an overestimation of our error budget following Eq.\ref{eq:cosmic_variance_bound}.

In order to derive the cosmic variance error contribution on the redshift distribution, we scale the CV by the number counts in redshift bins as predicted by the empirical likelihood framework (or alternatively by our deconvolution algorithm) using
\begin{equation}
    E[N_i] = N_{\rm tot} \, \pi_i^{\rm ML} \, ,
\end{equation}
where $\pi_i^{\rm ML}$ defines the maximum empirical likelihood estimate in redshift bin $i$ as discussed in \S~\ref{subsec:sample_pz_PhotZ}, and ($N_i$/$N_{\rm tot}$) denotes the (redshift bin $i$/total number of galaxies) in the tomographic bin.

Using the method of moments we can now estimate the parameters $\boldsymbol{\mu}$ and $\Sigma$ defined in Eq.~\eqref{eq:logistic_gp} as:
\begin{equation}
    \begin{split}
        \mu_i &= \log{\left(\frac{E[N_i]}{\sqrt{{\rm CV}_i^{2} + 1 }}\right)} \\
        \sigma_i^2 &= \log{\left({\rm CV_i}^2 + 1\right)}
    \end{split} \, ,
    \label{eq:method_moments_lgp}
\end{equation}
where the coefficient of variation is given as
\begin{equation}
    {\rm CV_i} = \sigma[N_i]\big/E[N_i] \, .
\end{equation}
This allows us to specify the logistic Gaussian process prior in Eq.~\eqref{eq:logistic_gp} defined in $N_{\rm bins}$ redshift bins for each of the four tomographic bins in our sample.

Given these definitions we can simplify the specification of the logistic Gaussian process on the parameters $\boldsymbol{\phi_\text{nz}}$ in Eq.~\eqref{eq:parametrization_pz} to
\begin{equation}
\begin{split}
\boldsymbol{s} &\sim \mathcal{N}(\boldsymbol{s} | \mu, \Sigma_{\rm CV}) \\
    \boldsymbol{\phi_\text{nz}} &:= \left\{\frac{\exp(s_i)}{\sum_j \exp(s_j)}  \ \bigg|  \ 0 < i < N_{\rm bins} \right\}   \, ,
    \end{split}
\end{equation}
where $\boldsymbol{\mu}$ and the diagonal matrix $\Sigma_{\rm CV}$ are defined in Eq.~\eqref{eq:method_moments_lgp}. The sampling of the $\boldsymbol{\phi_\text{nz}}$ parameters is expressed in terms of the variable $\mathbf{s}$ that follows a multivariate normal distribution.
This corresponds to the definition in Eq.~\eqref{eq:logGPsimple}.

\section{Marginalizing over the CAMIRA LRG photometric redshift error}
\label{sec:marg_camira_lrg}
In the following we describe the definition of the marginal likelihood that accounts for the photometric redshift error of the CAMIRA LRG (LRG) sample introduced in \S~\ref{subsec:camira_data}.
In this approach we treat the redshifts of each LRG as a latent variable. Since we do not have access to the likelihood of the photometric redshift method implemented in the CAMIRA method, we utilize the calibration dataset described in \S~\ref{subsec:spec_reference} to estimate a conditional distribution between the flux of the LRGs $\mathbf{f}_{\rm LRG}$ and their redshift $z_{\rm LRG}$. This is done by matching the LRG catalog and the specXphot calibration catalog and constructing a kernel based conditional density estimate. We can then marginalize the likelihood of spatial cross-correlations between the LRG and HSC photometric sample (phot) as:
\begin{equation}
\begin{split}
    &p(\mathbf{\hat{w}_{\rm LRG-PhotZ}} | \boldsymbol{\phi_\text{nz}}_{\rm PhotZ}, \mathbf{b_{\rm PhotZ}}, \mathbf{b_{\rm LRG}})  \\ &= \iint p(\mathbf{\hat{w}_{\rm LRG-PhotZ}} | \boldsymbol{\phi_\text{nz}}_{\rm PhotZ}, \mathbf{b_{\rm PhotZ}}, \mathbf{b_{\rm LRG}}, \mathbf{z_{\rm LRG}}, \mathbf{f}_{\rm LRG}) \\ &\times p(\mathbf{z_{\rm LRG}} | \mathbf{f}_{\rm LRG})  p(\mathbf{f}_{\rm LRG}) \, d\mathbf{f}_{\rm LRG} \, d\mathbf{z_{\rm LRG}}  \, ,
    \end{split}
\end{equation}
where $\mathbf{\hat{w}_{\rm LRG-PhotZ}}$ denotes the spatial cross-correlation measurements between the LRG and HSC phot catalogs, $\boldsymbol{\phi_\text{nz}}_{\rm PhotZ}$ denotes the $p_{\rm samp}(z)$ parameters of the HSC phot sample, and $(\mathbf{b_{\rm PhotZ}}$/$\mathbf{b_{\rm LRG}})$ is the galaxy-dark matter bias of the (HSC phot/CAMIRA LRG) sample. The left-hand-side defines the marginal likelihood introduced in Eq.~\eqref{eq:wx_like}. The term $p(\mathbf{z_{\rm LRG}} | \mathbf{f}_{\rm LRG})$ is the aforementioned conditional distribution of the LRGs' redshift given their flux. We also include the lensing weights for the HSC phot sample by weighting the pair counts used to construct the measurement $\mathbf{\hat{w}_{\rm LRG-PhotZ}}$ according to the prescription implemented in `The-Wizz' \citep{Morrison2016}.

Since the cross-correlation measurements don't vary much between realizations of LRG redshifts drawn from  $p(\mathbf{z_{\rm LRG}} | \mathbf{f}_{\rm LRG})$, we can evaluate this double integral using a Monte Carlo estimate:
\begin{equation}
    \begin{split}
    &\hat{p}(\mathbf{\hat{w}_{\rm LRG-PhotZ}} | \boldsymbol{\phi_\text{nz}}_{\rm PhotZ}, \mathbf{b_{\rm PhotZ}}, \mathbf{b_{\rm LRG}}) \\ &= \frac{1}{M} \sum_{(\mathbf{f_{\rm LRG}}, \mathbf{z_{\rm LRG}})} \left(p(\mathbf{\hat{w}_{\rm LRG-PhotZ}} | \boldsymbol{\phi_\text{nz}}_{\rm PhotZ}, \mathbf{b_{\rm PhotZ}}, \mathbf{b_{\rm LRG}}, \mathbf{z_{\rm LRG}}, \mathbf{f}_{\rm LRG}) \right) \, ,
    \end{split}
\end{equation}
where we sample $M$ sets $(\mathbf{f_{\rm LRG}}, \mathbf{z_{\rm LRG}})$ from the estimated joint distribution $\hat{p}(\mathbf{f_{\rm LRG}}, \mathbf{z_{\rm LRG}})$ by sampling sequentially as
\begin{equation}
    \begin{split}
        \mathbf{f_{\rm LRG}} &\sim \hat{p}(\mathbf{f_{\rm LRG}}) \\
        \mathbf{z_{\rm LRG}} &\sim \hat{p}(\mathbf{z_{\rm LRG}} | \mathbf{f}_{\rm LRG}) \, .
    \end{split}
\end{equation}
In this sampling scheme one has to recalculate the lensing-weighted pair-counts for each replication. This has the advantage that the scales and redshift bins can be consistently selected, but the disadvantage of high computational cost. However we iterate and verify that the variance in the integrand is moderate due to the small LRG photometric redshift error. Accordingly, we can use a small number of realizations ($M = 10$ in our case), which makes this a practical approach. We finally note that we speed up the construction of the conditional density estimate $\hat{p}(\mathbf{z_{\rm LRG}} | \mathbf{f}_{\rm LRG})$ by training directly on the residuals between the specXphot `true' redshifts in the training set $\mathbf{z_{\rm LRG}}$ and the estimated mean photometric redshift estimates $\mathbf{z}_{\rm phot, LRG}$, i.e. we construct $\hat{p}(\mathbf{z_{\rm LRG}} | \mathbf{z}_{\rm phot, LRG})$. 

While this potentially increases the variation in the resampled CAMIRA LRG redshifts, since we don't use the full information in the photometry as predictors, it allows us to train our error model efficiently on subsamples of LRG galaxies with very small variations between the conditional density function estimates due to the higher density of LRG training galaxies in the one-dimensional covariate $\mathbf{z}_{\rm phot, LRG}$.

\section{Discussion of Prior Choice}
\label{sec:PriorChoice}
In this appendix we discuss methodological differences between the logistic Gaussian Process as a prior over the $p_{\rm samp}(z)$ and the established alternative choice of the Dirichlet.

We have introduced the logistic Gaussian process as a prior distribution over $p_{\rm samp}(z)$ in \citet{2020MNRAS.491.4768R}, where we discuss several advantages in terms of characterizing the covariance between neighboring redshift bins. Furthermore, as explained in \S~\ref{sec:characterizing_cv_loggp} we can relate our choice of logistic Gaussian process prior to a lognormal model for the one-point density along the line-of-sight.

The Dirichlet distribution is an alternative prior that can be imposed over coefficients of finite basis function models like e.g. the histogram. It is a conjugate prior to the multinomial likelihood which is a significant advantage in designing sampling and inference schemes as, for example, demonstrated in the derivation of the variational inference scheme in \S~\ref{sec:derivation_inference_scheme}. In this context it is often applied as an uninformative prior over the histogram heights.

The Dirichlet distribution is related to a gamma distribution ${\rm Gamma}(\boldsymbol{\alpha}, 1)$ in a similar way as the logistic Gaussian Process to the lognormal model.
\begin{equation}
        \begin{split}
        \rho_i &\sim{\rm Gamma}(\boldsymbol{\alpha}, 1) \\
        \mathbf{N} &\sim {\rm MultNominal}(\boldsymbol{\phi_\text{nz}}, \overline{N})  \\
       \phi_{\rm nz, i} &= \frac{\rho_i}{\sum_{i = 0}^{N_{\rm bins}} \rho_i}
        \, ,
    \end{split}
\end{equation}
where the vector $\mathbf{N}$ denotes the galaxy counts drawn from the multinomial and $\overline{N}$ denotes the total number of galaxies. The vector $\boldsymbol{\phi_\text{nz}}$ would then be distributed according to a Dirichlet distribution with coefficients $\boldsymbol{\alpha}$.

Concentrating on the distribution of $\boldsymbol{\rho}$, that describes the expected number density of the point process along the line of sight, the logistic Gaussian process as the prior over the $p_{\rm samp}(z)$ implies a lognormal model, dependent on both a mean vector and covariance, whereas the choice of a Dirichlet distribution implies a one parameter Gamma distribution, dependent on the vector $\boldsymbol{\alpha}$. The limitation of the one-parameter Gamma distribution is that both the coefficient of variation and the average number density depend on the same parameter vector $\boldsymbol{\alpha}$. This means that we cannot parametrize a redshift-dependency in the coefficient of variation in the \referee{Dirichlet} model while leaving the mean histogram heights constant. We can however change the average coefficient of variation while leaving the mean constant as demonstrated in the following. Following \citet{Minka2000} we can reparametrize the Dirichlet as:
\begin{equation}
    \begin{split}
        s &= \sum_{k} \alpha_k \\
        \boldsymbol{m} &= \frac{\boldsymbol{\alpha}}{s} \, ,
    \end{split}
\end{equation}
where ($\boldsymbol{m}$/s) relates to the (mean/precision) of the Dirichlet distribution over the histogram heights. When the mean $\mathbf{m}$ is kept constant, one can modify the standard deviation of the Dirichlet distribution by scaling the precision $s$. \citet{2020arXiv200409542S} mention this aspect in their work in a slightly different context. We further note that typically the lognormal distribution can be adjusted to be close to the Gamma distribution.

In summary we use the logistic Gaussian Process model in this work as it allows a more flexible parametrization of uncertainty compared with the Dirichlet model.

% Don't change these lines
\bsp	% typesetting comment
\label{lastpage}
\end{document}